\documentclass[lettersize,journal]{IEEEtran}
\usepackage{amsmath,amsfonts}
\usepackage{algorithm}
\usepackage{algpseudocode}
\usepackage{array}
\usepackage{textcomp}
\usepackage{stfloats}
\usepackage{url}
\usepackage{soul}
\usepackage{verbatim}
\usepackage{graphicx}
\usepackage{xcolor}
\usepackage{quantikz}
\usepackage{multirow}
\usepackage{cite}
\usepackage{booktabs}
\usepackage{subcaption}
\newcommand{\INPUT}{\item[\textbf{Input:}]}
\newcommand{\OUTPUT}{\item[\textbf{Output:}]}
\begin{document}

\IEEEoverridecommandlockouts
\IEEEpubid{\makebox[\columnwidth]{This manuscript is a preprint currently under review at IEEE Transactions on Audio, Speech and Language Processing}}

\title{Quantifying Quanvolutional Neural Networks Robustness for Speech in Healthcare Applications}

\author{Ha Tran$^{1}$, \textit{Student Member, IEEE}, Bipasha Kashyap$^{1}$, \textit{Member, IEEE}, Pubudu N. Pathirana$^{1}$, \textit{Senior Member, IEEE}
\thanks{$^{1}$All authors are with the Networked Sensing \& Biomedical Engineering (NSBE) Research lab, School of Engineering, Deakin University, Waurn Ponds, Victoria, Australia (e-mail: thi.n.tran@deakin.edu.au, b.kashyap@deakin.edu.au, pubudu@deakin.edu.au}}%




\maketitle
\begin{abstract}
Speech-based machine learning systems are sensitive to noise, complicating reliable deployment in emotion recognition and voice pathology detection. We evaluate the robustness of a hybrid quantum machine learning model, quanvolutional neural networks (QNNs) against classical convolutional neural networks (CNNs) under four acoustic corruptions (Gaussian noise, pitch shift, temporal shift, and speed variation) in a clean-train/corrupted-test regime. Using AVFAD (voice pathology) and TESS (speech emotion), we compare three QNN models (Random, Basic, Strongly) to a simple CNN baseline (CNN-Base), ResNet-18 and VGG-16  using accuracy and corruption metrics (CE, mCE, RCE, RmCE), and analyze architectural factors (circuit complexity or depth, convergence) alongside per-emotion robustness. QNNs generally outperform the CNN-Base under pitch shift, temporal shift, and speed variation (up to 22\% lower CE/RCE at severe temporal shift), while the CNN-Base remains more resilient to Gaussian noise. Among quantum circuits, QNN-Basic achieves the best overall robustness on AVFAD, and QNN-Random performs strongest on TESS. Emotion-wise, \textit{fear} is most robust ($\approx$ 80–90\% accuracy under severe corruptions), \textit{neutral} can collapse under strong Gaussian noise ($\approx$5.5\% accuracy), and \textit{happy} is most vulnerable to pitch, temporal, and speed distortions. QNNs also converge up to six times faster than the CNN-Base. To our knowledge, this is a systematic study of QNN robustness for speech under common non-adversarial acoustic corruptions, indicating that shallow entangling quantum front-ends can improve noise resilience while sensitivity to additive noise remains a challenge.
\end{abstract}


\section{Introduction}
The power of machine learning (ML) methods has allowed for monumental advancements in many tasks such as automatic speech recognition \cite{yu2016automatic, ahlawat2025automatic}, speaker identification \cite{wang2020data}, speech emotion recognition \cite{meng2019speech}, and voice pathology detection \cite{harar2017voice, mohammed2020voice}. However, unlike humans, such ML methods can be highly sensitive to noisy data. This sensitivity is acute in real-world deployments where acoustic signals face additive, spectral, temporal, and channel corruptions. In everyday speech communication, acoustic signals are exposed to a wide variety of corruptions, including background noise, reverberation, channel distortions, and speaker-induced variability such as pitch shifts, tempo changes, and prosodic fluctuations. These corruptions are one of the key challenges in developing effective ML frameworks for speech tasks. This is especially true in medical applications that aim to extract health-related information and assist diagnosis such as speech affect (emotion) recognition and voice pathology detection. Speech affect recognition seeks to identify a speaker's emotional sate, and voice pathology detection focuses on detecting abnormalities in the voice from vocal cues, using signal processing and ML methods. Both applications often contend with noisy data and can degrade under inadequate models. Consequently, these systems require ML frameworks that are not only accurate but also resilient to noise.

Many studies have analysed the robustness of ML against noisy signals, and proposed techniques to mitigate the effect of noise. Many techniques aim to engineer features that are robust to noise, such as signal-level noise reduction \cite{boll2003suppression, kamath2002multi, koo1989filtering}.  
Instead of focusing on engineering robust features, in this study, we aim to improve robustness by designing suitable model architecture. Specifically, rather than training the ML models on noisy data with robust features, we train the models on clean data and evaluate their robustness on unseen, corrupted data. This method allows us to improve the resilience and consistency of the models when evaluated on real-world data, corrupted by various types of noise. This is beneficial in applications that work with diverse, noisy data, such as medical data, and require high reliability. Researchers have shown that the robustness of deep learning models come from their ability of learning invariant representations. For example, Convolutional Neural Networks (CNNs), and Recurrent Neural Networks (RNNs) have demonstrated good performance on noisy data in recognition tasks that deal with noisy environments \cite{qian2016very, tan2018adaptive}. Residual Networks (ResNet) employ the residual connections that enable the model to focus on relevant spectral-temporal patterns, reduce the vanishing gradient problem, and suppress irrelevant noise \cite{ao2024used, zhou2021resnext}. Architectures such as VGG-like CNNs further extend this robustness by leveraging hierarchical feature extraction \cite{hamsa2023speaker}.
\IEEEpubidadjcol

Recently, quantum machine learning (QML) has emerged as a promising technique that combines the flexibility and learning power of ML, robustness of quantum algorithms, and efficiency of quantum hardware. QML aims to enhance security, computational efficiency, and accuracy of ML approaches in many speech applications. Studies have shown that QML approaches can achieve better performance than classical ML approaches in speech recognition \cite{yang2021decentralizing, sridevi2023quantum}, and speech emotion recognition \cite{balachandran2025advanced, soltani2025quantum, chen2025consensus}. Researchers have also demonstrated that QML can have superior performance, albeit on limited speech datasets, which in itself is a significant challenge in many applications such as healthcare \cite{sridevi2023quantum}. However, the application of QML in the speech domain is still very limited compared to classical ML. Furthermore, in applications that require good resilience against noisy data such as speech, the robustness of QML is yet to be explored. The lack of an analysis of the robustness of QML in the speech domain is indeed the motivation for our study.

In this study, we present a robustness assessment framework for QML models in the area of speech. Our contributions in particular are as follows:
\begin{itemize}
    \item Robustness evaluation on two speech tasks: We systematically assess QNNs on voice-pathology detection (AVFAD) and speech-emotion recognition (TESS) in a clean-train/corrupted-test setting. For emotion recognition, we further analyse class-wise robustness to identify which emotions remain stable and which degrade most under each corruption.
    \item Quantum–classical comparison under four corruption types: We compare QNNs and CNNs across six severities of four non-adversarial acoustic corruptions (Gaussian noise, pitch shift, temporal shift, and speed variation) using accuracy and corruption metrics. 
    \item Effect of quantum circuit design on robustness:  We quantify how the quantum circuit and circuit depth influence robustness, reporting architecture-specific trade-offs and identifying configurations that achieve the best robustness profiles on each dataset.
\end{itemize}

The rest of this study is organized as follows. In Section \ref{sec: related work}, we present related studies on QML algorithms and their robustness. The next Section \ref{sec: background} explains the fundamental concept of quantum computing and three types of quantum circuits. Section \ref{sec: methodology} presents the speech signal processing, model architecture. Section \ref{sec: experiments} describes the datasets, the four types of corruption, training, and testing procedures. Section \ref{sec: results and discussion} demonstrates the results, and provides a discussion on the classification accuracy and robustness. Finally, Section \ref{sec: conclusion} gives a summary of the key findings.

\begin{figure*}[htbp]
    \centering
    \includegraphics[width=0.93\textwidth]{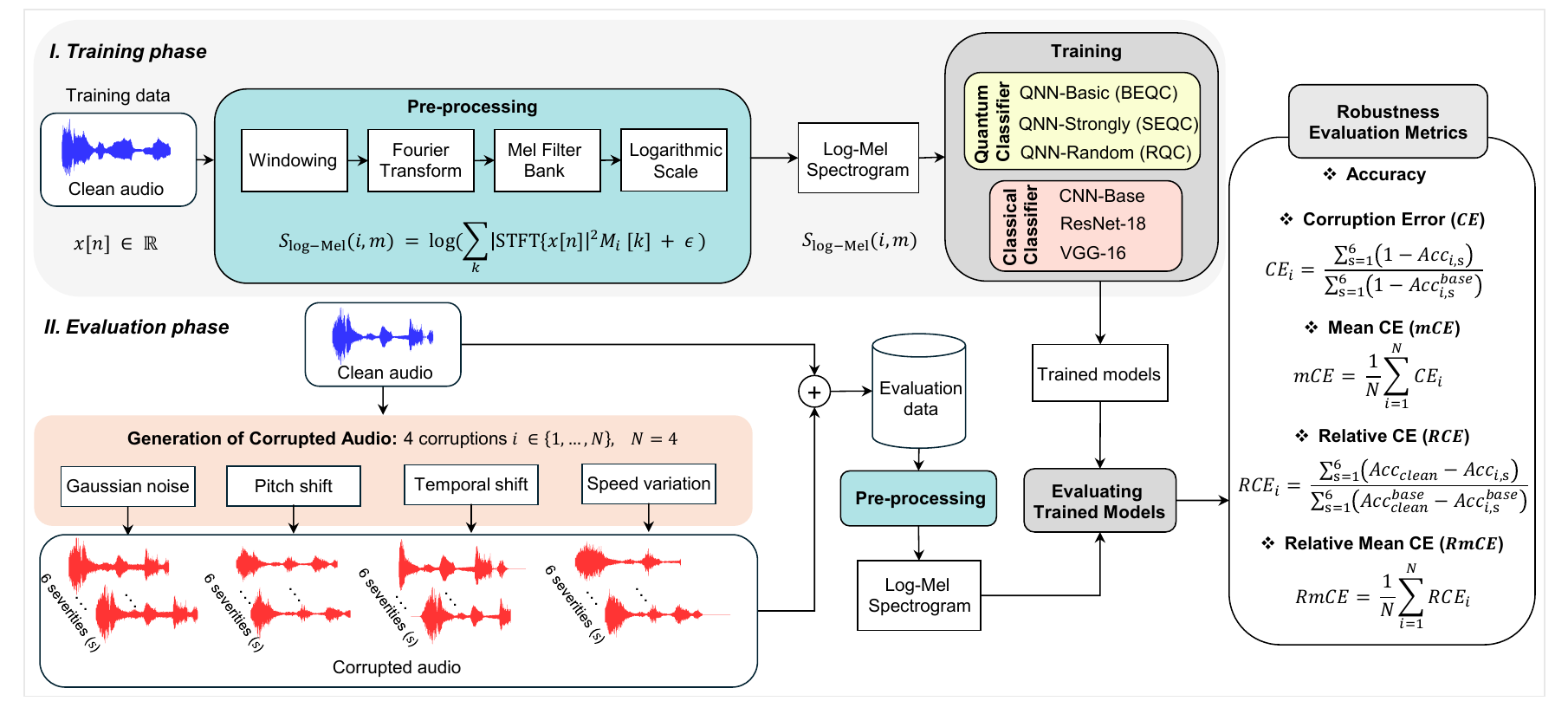}
    \caption{The proposed Quantum-Classical robustness evaluation framework for speech classification. Clean audio is converted into log-Mel spectrograms and used to train QNN models (QNN-Basic, QNN-Strongly, QNN-Random) and classical CNNs (CNN-Base, ResNet-18, VGG-16) classifiers. Model robustness is evaluated on corrupted audio (Gaussian noise, pitch shift, temporal shift, speed variation) using metrics such as CE, mCE, RCE, and RmCE.}
    \vspace{-15pt}
    \label{fig:framework}
\end{figure*}

\section{Related Work}
\label{sec: related work}

\subsection{The efficiency of Quantum Machine Learning}

There are many studies on popular QML algorithms that have demonstrated good results on speech processing, such as quantum support vector machine \cite{sridevi2023quantum}, quantum neural networks \cite{esposito2022quantum}, quantum convolutional neural network \cite{balachandran2025advanced}. To combine the expressiveness, robustness of CNNs, and the efficiency of QML, hybrid QML approaches have been explored. Hybrid approaches utilize the strength of both classical model and QML model to enhance their performance. A notable example is the hybrid Quanvoltional Neural Network (QNN) model \cite{henderson2020quanvolutional}. By leveraging the properties of quantum mechanics in convolutional transformation, QNN aims to further advance the extraction of feature and the overall processing of information for ML applications \cite{yang2021decentralizing, ceschini2025hybrid, ahalya2023ranet}. Thus, there is a demand for investigating the robustness of QNN in various ML applications.  
\vspace{-1em}
\subsection{Robustness of Quantum Machine Learning}
Noisy Intermediate-Scale Quantum (NISQ) devices \cite{preskill2018quantum} currently suffer from noise caused by the devices, such as decoherence, gate errors, and interference from the environment. Many studies have focused on mitigating this type of noise in NISQ devices using quantum framework \cite{guan2021robustness, li2025qmlsc}. One study \cite{ahmed2025comparative} investigated the robustness of QNN against quantum noise on image data and their results shown that QNN achieved higher robustness compared other QML model. This demonstrates the potential of QNNs for noise resilience. In this study, however, we focus on noise that disturbs the input signals, instead of the noise generated by quantum devices.

We first consider adversarial attacks, which deliberately perturb input data. Adversarial attacks refer to small, intentional perturbation added to input signals. This type of noise is often imperceptible by human; however, it can cause ML models to make wrong predictions. Existing research has explored the robustness of quantum models against adversarial perturbations, though most work has focused on image data. An example is \cite{lu2020quantum}, which showed that QML is also susceptible to specific types of adversarial attacks, similar to classical ML. In \cite{west2023benchmarking}, the authors found that variational classifiers built upon QML can learn features undetectable to classical neural networks. The additional features can increase the robustness of quantum classifier, mitigating the effects of adversarial attacks. However, their approach are very computationally expensive, which require $10^3-10^4$ quantum gates, exceeding current quantum hardware capabilities. In \cite{el2024advqunn}, the robustness of QNN was evaluated against three types of adversarial attacks. It was demonstrated that increasing the quantum entanglement property of QNNs can improve their robustness compared to classical CNNs. Despite the many studies of adversarial attacks in the imaging domain, research on their effects in speech processing applications of QML is still limited. For example, \cite{wang2023enhancing} investigated QNNs for protecting the privacy of audio-visual speech under white-box and black-box attacks, but only analysed a single quantum circuit configuration. The robustness of other quantum circuit architectures in speech applications remains largely unexplored.

Beyond adversarial perturbations, another type of noise that affect input signals is random noise generated by real-world variations. This can include background noise, differences among the speakers, and variations in processing. \cite{quan2025quantum} proposed transformers inspired by quantum properties to classify acoustic scenes. Across a wide range of acoustic environments, the proposed transformers showed better robustness and accuracy compared to classical models when evaluated on speech data perturbed by white Gaussian noise. Unlike acoustic scene classification, our study focuses on natural voice corruptions affecting human speech. To the best of our knowledge, this is the first study to investigate the robustness of QNNs against natural acoustic corruptions in speech classification tasks. The goal is to establish an experimental baseline for comparing the robustness of quantum and classical ML methods under realistic input degradations.

\vspace{-1em}
\subsection{Research Gap and Motivation}
Existing studies lack (1) systematic comparison of quantum circuit architectures for speech processing and (2) comprehensive analysis of quantum model robustness under natural acoustic corruptions. This motivates evaluating how circuit design influences QNN robustness in realistic speech conditions. QNNs offer advantages for high-dimensional speech data through spatially local quantum transformations and shallow circuits suitable for NISQ devices. Their hybrid architecture allows seamless integration with classical deep learning while enabling quantum-enhanced feature extraction. Inspired by \cite{hendrycks2019benchmarking}, which benchmarked robustness in image models, this work investigates QNN robustness against four common acoustic corruptions: Gaussian noise, pitch shift, temporal shift, and speed variation.

\section{Background}
QML is the combination of quantum computing and ML. Therefore, QML uses properties of quantum such as superposition, entanglement to enhance ML. This section will provide the fundamentals of quantum computing and three types of quantum circuits. 
\label{sec: background}
\vspace{-1em}
\subsection{Quantum Computing Fundamentals}
Quantum computing exploits the principles of quantum mechanics to perform computations \cite{nielsen2010quantum}. Analogous to the bit in classical computation, a quantum bit (qubit) is the fundamental concept of quantum computation. However, it can exist in a linear combination of 0 and 1. This superposition forms the foundation of quantum speed-ups in algorithms such as Grover’s search \cite{grover1996fast}. In QML, superposition enables embedding classical data into high-dimensional Hilbert space, which can classify pattern easier like Support Vector Machine \cite{schuld2019quantum}. Quantum gates are the basic operations on qubits and serve as the fundamental blocks of quantum circuits, analogous to logic gates in classical computing. Quantum gates are categorized into two main types: single-qubit gates and multi-qubit gates. Single-qubit gates operate on individual qubits such as Hadamard gate ($H$), and rotation gates ($R_x(\theta)$, $R_y(\theta)$, $R_z(\theta)$). In contrast, multi-qubit gates operating on two or more qubits simultaneously include the controlled-NOT (CNOT), controlled-Z (CZ), SWAP, and Toffoli (T) gates. Some multi-qubit gates such as CNOT, CZ enable correlations between qubits to create entanglement property. In QML, entanglement increases the expressive capacity of quantum circuits by creating correlations among features \cite{el2024advqunn}. A quantum circuit, which is a programmable sequence of quantum gates acting on qubits, process data within the Hilbert space, exploit superposition and entanglement properties. 

\subsection{Quantum Circuit Architectures}
\label{sec:quantum_circuits}
\begin{figure}[!t]
    \centering
    \begin{subfigure}{0.35\textwidth}
        \centering
        \includegraphics[width=\linewidth]{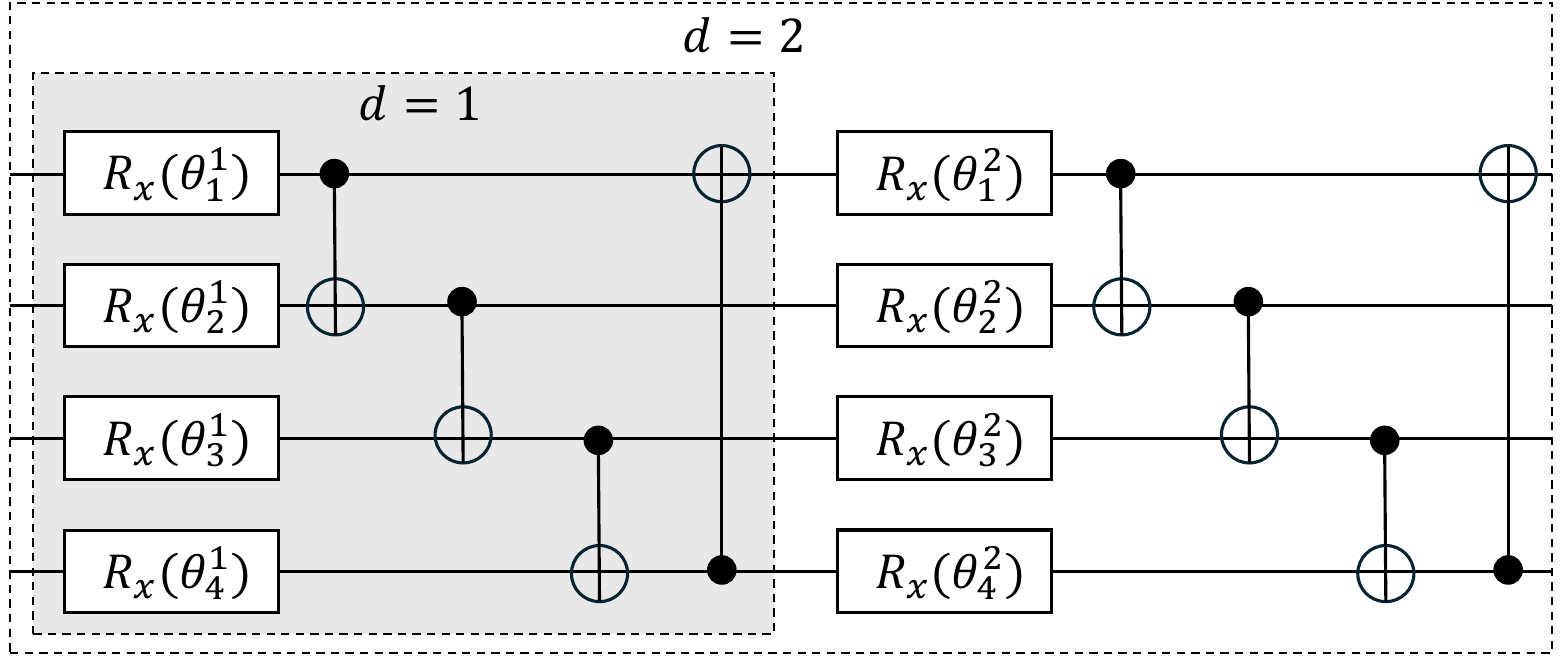}
        \caption{BEQC. The same circuit layer is repeated as the depth increases.}
        \label{fig:basic_quantum_circuit}
    \end{subfigure}
    \hfill
    \begin{subfigure}{0.45\textwidth}
        \centering
        \includegraphics[width=\linewidth]{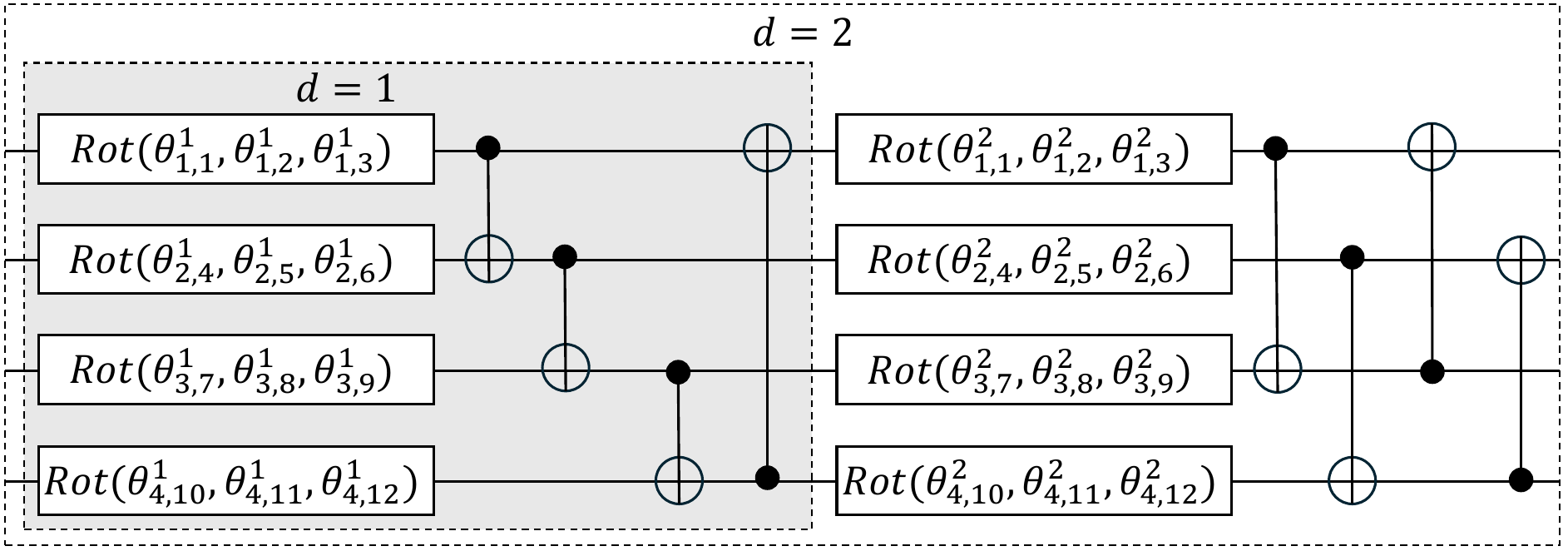}
        \caption{SEQC. The rotation blocks ($\mathrm{Rot}$), including a sequence of $R_z, R_y$, and $R_z$, are reused, while the CNOT connections vary across circuit layers.}
        \label{fig:strongly_quantum_circuit}
    \end{subfigure}
    \hfill
    \begin{subfigure}{0.48\textwidth}
        \centering
        \includegraphics[width=\linewidth]{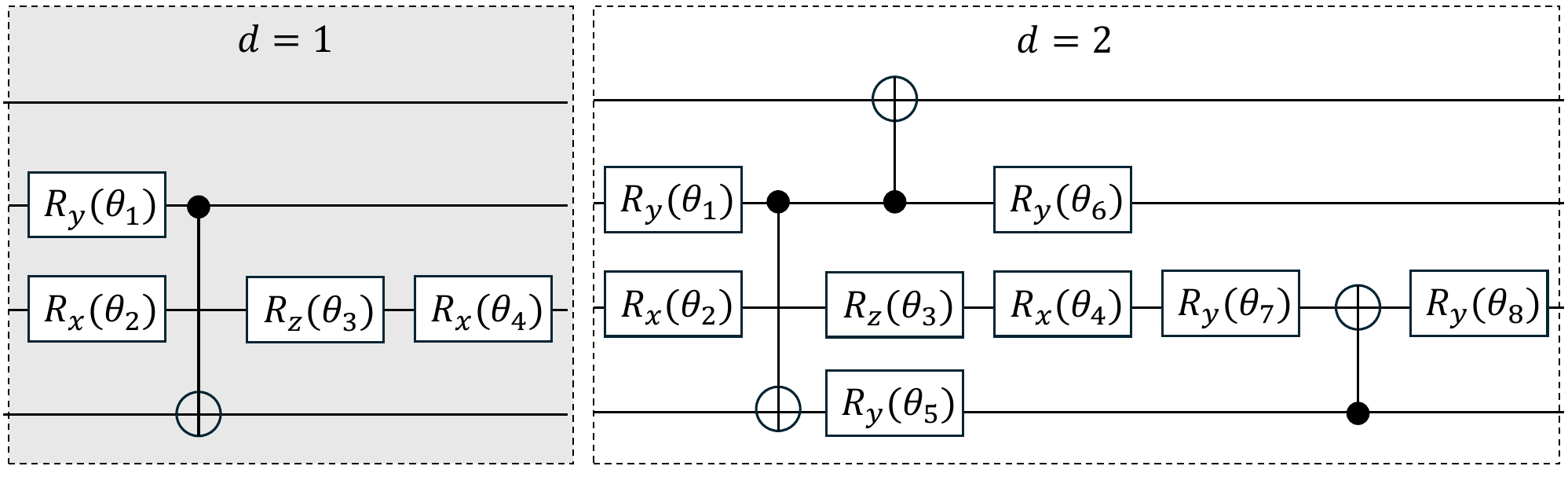}
        \caption{RQC. Each additional layer is randomly sampled, resulting in distinct circuit structures at higher depths.}
        \label{fig:random_quantum_circuit}
    \end{subfigure}

    \caption{Examples of three quantum circuit architectures for $n=4$ qubits: 
    (a) BEQC, 
    (b) SEQC, and 
    (c) RQC, 
    each illustrated for circuit depths $d=1$ and $d=2$.}
    \vspace{-6pt}
    \label{fig:all_quantum_circuits}
\end{figure}

This section examines three representative quantum circuit templates from Pennylane \cite{bergholm2018pennylane} frequently used in QNN studies \cite{el2024advqunn, ahmed2025comparative, yang2021decentralizing}: the Basic Entanglement (BEQC), the Strongly Entanglement (SEQC), and the Random Quantum Circuit (RQC). Each architecture exhibits a distinct structure and connectivity pattern, influencing how quantum information is transformed across layers. Rather than proposing novel architectures, this work focuses on analysing their behaviour and robustness across different circuit depths ($d$), defined as the number of repeated circuit layers.
\subsubsection{Basic Entanglement Quantum Circuit (BEQC)}
The BEQC follows a structured design that integrates parametrized single-qubit rotations ($R_x$) with a closed-chain configuration of CNOT gates for $n$ qubits, as illustrated in Fig.~\ref{fig:basic_quantum_circuit}.  
The entangling block is defined by the unitary operator
\begin{equation}
E^{(i)}_{\mathrm{basic}} \;=\;
\mathrm{CNOT}_{n \rightarrow 1}
\Bigg( \prod_{q=1}^{n-1} \mathrm{CNOT}_{q \rightarrow q+1} \Bigg),
\label{eq:entangler_basic}
\end{equation}
forming a ring topology that ensures cyclic nearest-neighbour entanglement.  
A depth-$d$ BEQC is obtained by stacking the same layer $d$ times:
\begin{equation}
U_{\text{basic}}^{(d)}(\boldsymbol{\theta}) \;=\;
\prod_{i=1}^{d}
\Bigg[
E^{(i)}_{\mathrm{basic}}
\prod_{q=1}^n R_x^{(q)}\!\big(\theta^{(i)}_q\big)
\Bigg],
\label{eq:basic_circuit}
\end{equation}
where $\boldsymbol{\theta} = \{\boldsymbol{\theta}^{(1)}, \dots, \boldsymbol{\theta}^{(d)}\}$ denotes the complete set of rotation parameters across all layers.

\subsubsection{Strongly Entanglement Quantum Circuit (SEQC)}
The SEQC architecture maximizes inter-qubit correlations by employing dense entanglement and alternating gate connectivity between layers. Each layer first applies a sequence of three parametrized single-qubit rotations ($R_z, R_y, R_z$) to $n$ qubit, collectively denoted as $\mathrm{Rot}$:
\begin{equation}
U_{\mathrm{Rot}}^{(i)}(\boldsymbol{\theta}^{(i)}) \;=\;
\prod_{q=1}^{n}
R_z^{(q)}(\theta_{q,1}^{(i)})\,
R_y^{(q)}(\theta_{q,2}^{(i)})\,
R_z^{(q)}(\theta_{q,3}^{(i)}),
\label{eq:strong-urot}
\end{equation}
followed by an entangling block that alternates the CNOT direction across layers:
\begin{equation}
E^{(i)}_{\mathrm{strongly}} \;=\;
\prod_{q=1}^{n} 
\mathrm{CNOT}_{\,q \rightarrow \big((q+i-1) \bmod n\big) + 1}.
\label{eq:entangler_strongly}
\end{equation}
The resulting depth-$d$ circuit is expressed as
\begin{equation}
U_{\mathrm{strongly}}^{(d)}(\boldsymbol{\theta}) \;=\;
\prod_{i=1}^{d} \Big[\, E^{(i)}_{\mathrm{strongly}} \; U_{\mathrm{Rot}}^{(i)}(\boldsymbol{\theta}^{(i)}) \,\Big].
\label{eq:strong-circuit}
\end{equation}
The alternating topology allows all qubits to interact across layers, enhancing the circuit’s ability to capture complex multi-qubit relationships.

\subsubsection{Random Quantum Circuit (RQC)}
The RQC is constructed by sampling unitary operations and arranging them into randomly parametrized layers. Each layer typically combines single-qubit rotation gates with randomly selected multi-qubit CNOT gates. In this study, a single RQC configuration is fixed across 10 random seeds to isolate the influence of input randomness arising from data corruption, rather than randomness in the circuit topology itself.
Unlike the structured BEQCs, RQCs do not reuse the same configuration as depth increases. For instance, at $d=1$, the circuit corresponds to the layout in Fig.~\ref{fig:random_quantum_circuit}, whereas at $d=2$, an additional randomly sampled layer is appended instead of repeating the previous structure.  
To facilitate fair comparison, the ratio of CNOT to rotation gates in the RQC is fixed at 0.3/0.7, resulting in approximately 1.71$d$ CNOT gates, given that the number of rotation gates in the RQC is 4$d$. This configuration yields a lower entanglement density compared to the BEQC and SEQC, which use more two-qubit connections and therefore achieve stronger coupling between qubits. Since RQCs are inherently stochastic, their analytical unitary forms are not explicitly represented; instead, comparative results are reported in terms of the gate composition (see Table~\ref{tab:circuit_comparison}).

\begin{table}[]
\centering
\caption{Comparison of three quantum circuits (BEQC, SEQC, RQC) in terms of the rotation gates, the number of rotation gates ($N_{Rotation}$), and the number of CNOT gates ($N_{CNOT}$) with $d$ is the depth of quantum circuit.}
\label{tab:circuit_comparison}
\resizebox{\columnwidth}{!}{%
\begin{tabular}{cccc}
\hline
Quantum Circuit & Rotation Gates  & $N_{Rotation}$ & $N_{CNOT}$  \\ \hline
BEQC            & $R_x$           & 4$d$             & 4$d$          \\
SEQC            & $R_z, R_y$      & 12$d$            & 4$d$          \\
RQC             & $R_x, R_y, R_z$ & 4$d$             & $\sim$1.71$d$ \\ \hline
\end{tabular}%
}
\vspace{-2em}
\end{table}

\section{Methodology}
\label{sec: methodology}
\subsection{Processing Speech}
\label{subsection: speech process}
In audio processing, several data representations can be used depending on the model architecture and application domain. Audio can be fed to a network either as its raw waveform (a 1D temporal signal) or as a 2D time–frequency representation, such as a Mel-spectrogram, log-Mel spectrogram, or Mel-Frequency Cepstral Coefficients (MFCCs). In most recent studies, both QNNs and CNNs tend to rely on 2D acoustic features, as they more effectively capture temporal and spectral dependencies. Among these representations, the log-Mel spectrogram has consistently demonstrated strong performance in tasks like speech recognition and voice pathology classification \cite{yang2021decentralizing, meng2019speech, meghanani2021exploration, bhangale2020speech}. Therefore, this study adopts log-Mel spectrograms as the input representation for all models. 

The log-Mel spectrogram converts an audio waveform $x[n] \in \mathbb{R}$ into a two-dimensional time–frequency representation that captures perceptually relevant spectral information. As illustrated in Fig.~\ref{fig:framework}, the waveform is first divided into overlapping frames using a windowing function to preserve local continuity. Each frame is then transformed into the frequency domain via the Short-Time Fourier Transform (STFT), producing a time-varying spectrum. The spectral energy is passed through a Mel filter bank and converted to the logarithmic scale according to
\[
S_{\mathrm{log\text{-}Mel}}(i,m) = \log \left( \sum_k | \mathrm{STFT}\{x[n]\} |^2 M_i[k] + \epsilon \right),
\]
where $M_i[k]$ denotes the $i^{th}$ Mel filter and $\epsilon$ is a small constant for numerical instability. In this study, log-Mel spectrograms are computed using the \texttt{Librosa} library with an FFT size of 512 samples, a hop length of 128 samples, a window length of 25~ms, 40 Mel filters, and $\epsilon = 10^{-10}$. The resulting spectrograms are normalized and resized to fixed dimensions of $40\times128$ (height$\times$width), serving as standardized 2-D inputs for the QNN architectures described in the following sections. 
\vspace{-1em}
\subsection{Quanvolutional Neural Network}
QNNs are the hybrid quantum-classical neural networks that use a quantum transformation layer as a convolutional layer \cite{henderson2020quanvolutional}. In a classical CNN, the convolutional layer applies filters to extract feature from input data. Similarly, the quantum convolutional (quanvolutional) layer replaces these filters with a quantum circuit. Therefore, in QNN the first convolutional layer is changed by a quanvolutional layer and the rest of the architecture remains classical. In this study, we evaluate three QNN configurations: QNN-Basic, QNN-Strongly, and QNN-Random, which vary in quantum circuit transformations and depth to assess how circuit complexity affects model robustness. 

\subsubsection{\textbf{Quanvolutional Layer}}
\label{subsub:Quanv layer}
The quanvolutional layer employs the quantum circuit to transform patches of the log-Mel spectrogram. The spectrogram is divided into small local regions and processed by the quanvolutional layer separately. In this study, a 2 $\times$ 2 patch of the input is selected to map to four qubits. Previous studies have shown that the \(2 \times 2\) kernel achieves a favourable trade-off between representational power and computational feasibility compared to alternative kernel such as \(1 \times 1\) or \(3 \times 3\) \cite{yang2021decentralizing, ceschini2025hybrid}. Moreover, this configuration is aligned with the capabilities of current NISQ devices, which makes it a practical choice for implementation. The quanvolutional layer consists of three main components: encoding ($S$), where classical data is embedded into quantum states; quantum circuit transformation ($U$), where unitary operations are applied to capture high-dimensional feature relationships; and decoding ($\mathcal{M}$), where measurements are performed to obtain classical feature maps. The resulting log-Mel feature representations after processing through the quanvolutional layer of the QNN and the convolutional layer of the CNN-Base are illustrated in Fig. 9 of the Supplementary Material.

\paragraph{Encoding}
Classical input patches, given by \( \mathbf{x} = [x_1, x_2, \ldots, x_n]^{\mathsf{T}} \in \mathbb{R}^n \) are encoded into quantum states within an \(n\)-qubit Hilbert space \( \mathcal{H} \) using an encoding function \( S : \mathbb{R}^n \rightarrow \mathcal{H} \). Each component \(x_i\) is the \(i^{\text{th}}\) element of the patch \(\mathbf{x}\) and is assigned to a single qubit; thus, a \(2 \times 2\) region of the log-Mel spectrogram contains four values and is represented using four qubits \((n = 4)\) initialized in the ground state \(|0\rangle\). Using angle encoding, each normalized pixel \(x_i \in [0,1]\) determines the rotation angle of a single-qubit gate \(R_y\), scaled by \(\pi\) to ensure a physically valid parameter range. This encoding scheme preserves the local spatial structure of the input patches and is compatible with current NISQ devices. Scaling by \(\pi\) ensures that the rotations span the Bloch hemisphere \([0,\pi]\), yielding non-redundant qubit states within the quantum feature space. The complete encoding is defined as
\[
|\phi(\mathbf{x})\rangle = S(\mathbf{x}) = \bigotimes_{i=1}^{n} R_y(\pi x_i)\,|0\rangle,
\label{eq:encoding}
\]
where \( \bigotimes \) denotes the tensor product over the \(n\) single-qubit states.

\paragraph{Quantum circuit transformation}
The encoded quantum state is then transformed by a quantum circuit represented by a unitary operator \( U \), which consists of single and multi-qubit gates. In this work, we analyse the impact of circuit complexity on the robustness of the quantum model for audio classification by evaluating three quantum architectures: BEQC, SEQC, and RQC as described in Section~\ref{sec:quantum_circuits}, corresponding to the QNN-Basic, QNN-Strongly and QNN-Random configurations. Furthermore, the circuit depth (\( d \)) is varied from 1 to 50 layers to investigate the effect of increasing quantum transformation depth. This process transforms the encoded patch into a richer quantum representation (\( |\psi(\boldsymbol{\theta}, \mathbf{x})\rangle \)) through superposition and entanglement, is expressed by
\begin{equation}
|\psi(\boldsymbol{\theta}, \mathbf{x})\rangle = U^{(d)}(\boldsymbol{\theta}) |\phi(\mathbf{x})\rangle,
\end{equation}
where \( \boldsymbol{\theta} \) denotes a fixed set of quantum parameters. Prior studies have shown that non-trainable quanvolutional layers can perform comparably to trainable ones~\cite{el2024advqunn, ceschini2025hybrid, henderson2020quanvolutional}. Based on this studies, we keep the quantum parameters fixed in our quanvolutional layer to minimize computational cost while preserving effectiveness. The trainable variant can be explored in future work.

\paragraph{Decoding}
Decoding (or measurement, \( \mathcal{M} \)) extracts classical information from the quantum state after circuit transformation \( U \). In this work, the Pauli-\( Z \) observable is applied to each qubit, producing expectation values that represent the average outcomes over many projective measurements. We use the \texttt{shots=None} configuration in the PennyLane backend, which provides analytical expectation values without the statistical noise from finite sampling~\cite{ceschini2025hybrid}. Four expectation values, denoted \( \mathbf{f}_{\boldsymbol{\theta}}(\mathbf{x}) \), correspond to four qubits, each ranging within \([-1, 1]\), as defined in Eq.~\ref{eq:measurement}. These values are assigned to four output channels (ch.1–4) in Fig.~\ref{fig:Quanvolutional layer}, where each \( 2 \times 2 \) input patch is transformed into a multi-channel feature representation, which is the output of the quanvolutional layer.
\begin{equation}
\mathbf{f}_{\boldsymbol{\theta}}(\mathbf{x}) = \langle \mathcal{M} \rangle_{\mathbf{x}, \boldsymbol{\theta}} = \langle \psi(\mathbf{x}, \boldsymbol{\theta}) | \sigma_z | \psi(\mathbf{x}, \boldsymbol{\theta}) \rangle.
\label{eq:measurement}
\end{equation}
\vspace{-1.5em}

\subsubsection{\textbf{Quanvolutional network}}
\begin{figure}[htbp]
    \centering
    \includegraphics[width=\columnwidth]{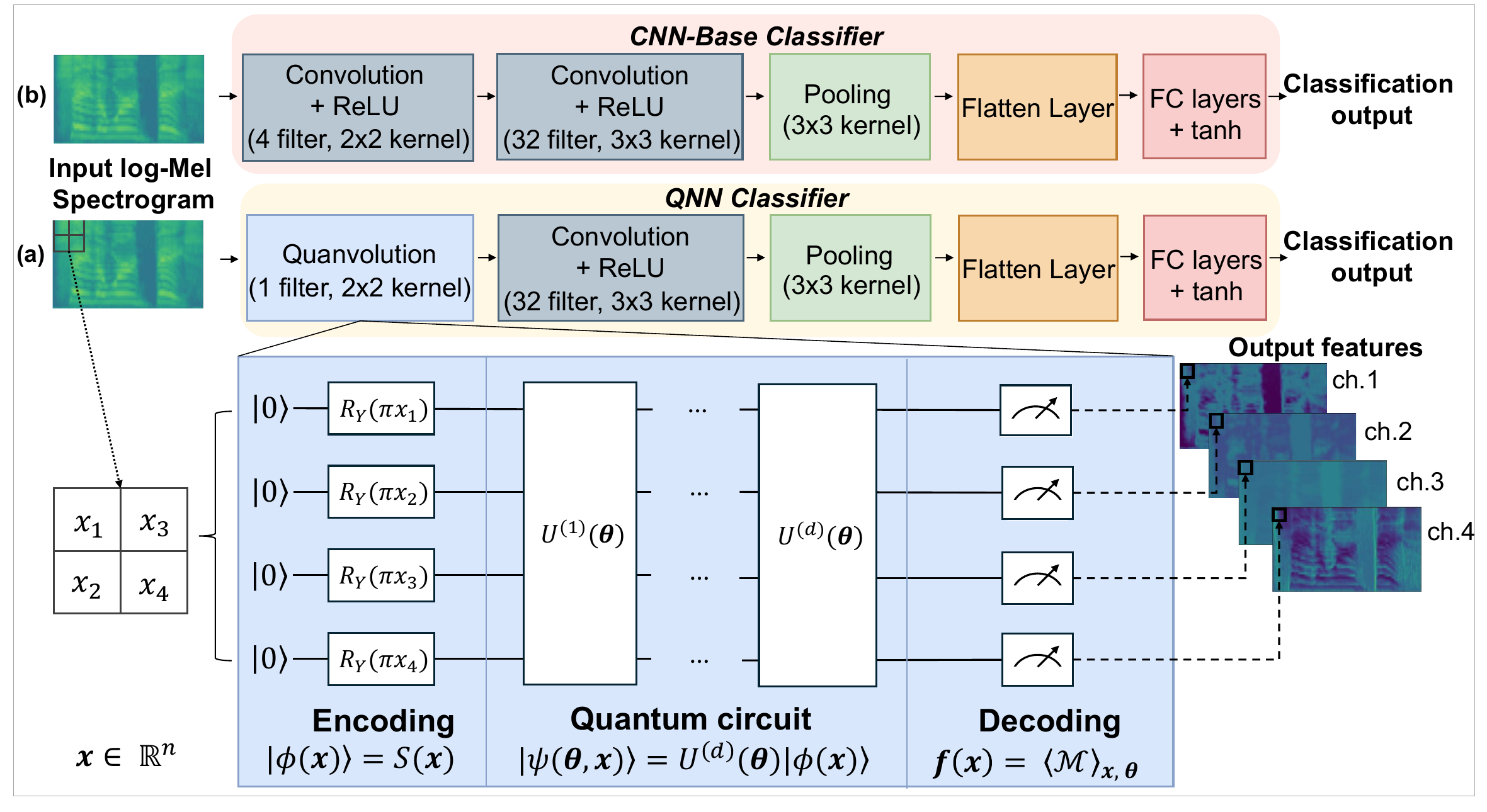}
    \caption{Architectures of (a) QNN and (b) CNN-Base classifiers. In (a), the quanvolutional layer includes encoding, quantum circuit, and decoding stages, where inputs are encoded by rotation gates ($R_y$), processed, and decoded into four classical channels (ch.1--ch.4). These channels are then passed through convolutional, pooling, flatten, and fully connected layers. The main difference lies in the first layer: QNN uses a quanvolutional layer, while CNN-Base uses a classical convolutional layer.}
    \vspace{-10pt}
    \label{fig:Quanvolutional layer}
\end{figure}

Similar to the classical convolutional layer, the quanvolutional layer can be easy integrated into a wide range of neural network architectures. For instance, this layer can be followed by a simple fully connected classical layers in \cite{el2024advqunn, ceschini2025hybrid}, or by a sequential layer of the convolutional layers, the pooling layers, and the fully connected layers in \cite{henderson2020quanvolutional}. In this study, we aim to leverage the complementary strengths of quantum and classical convolutional layers. Accordingly, our QNN architecture comprises a quanvolutional layer, a classical convolutional layer, a pooling layer, a flatten layer, and two fully connected layers (FC), as shown in Fig. \ref{fig:Quanvolutional layer}. The quanvolutional layer contains a single filter with a kernel size of $2 \times 2$. The subsequent convolutional layer employs a ReLU activation function, a $3 \times 3$ kernel size, and 32 filters. The pooling layer uses a $3 \times 3$ kernel, reducing the spatial dimension by a factor of three. After flattening, the feature maps are passed through the fully connected (FC) block, where the first fully connected layer consists of 64 hidden units with a tanh activation, and the second fully connected layer serves as the output layer, containing a number of units equal to the number of classes (labels). 

\subsection{Benchmark CNN models}
The primary objective of this study is to evaluate the robustness of QNNs in comparison with their classical counterparts. The main benchmark is a simple CNN baseline (CNN-Base) that mirrors the QNN architecture, except that the first quanvolutional layer is replaced by a conventional convolutional layer. This convolutional layer employed a ReLU activation function, a 2$\times$2 kernel, and four filters, as shown in Fig. \ref{fig:Quanvolutional layer}. This design ensures a fair, one-to-one comparison, isolating the contribution of quantum feature extraction to model robustness.

To further contextualize the results, two state-of-the-art classical architectures, ResNet-18 \cite{he2016deep} and VGG-16 \cite{simonyan2014very}, are also included as additional baselines representing advanced CNN models widely used in speech processing. Both models are employed as pre-trained networks initialized with ImageNet weights and fine-tuned for speech classification. Their first convolutional layers are modified to accept single channel log-Mel spectrograms, and their final classification layers are replaced with new linear layers producing the corresponding output classes. In addition, VGG-16 incorporates an adaptive average pooling layer to standardize the feature map size before the classifier.
These adaptations allow both pre-trained architectures to process 2D acoustic representations, enabling an efficiency comparison of QNNs, CNN-Base, and deeper pre-trained models in terms of robustness under various acoustic corruptions.
\section{Experiments}
\label{sec: experiments}
\subsection{Datasets}
In our experiments, we employ two audio datasets: one for voice pathology assessment and one for speech emotion recognition.  

\begin{itemize}
\item Advanced Voice Function Assessment Dataset (AVFAD) \cite{jesus2017advanced} includes recordings from 709 participants (346 with voice disorders and 363 healthy controls). It contains sustained vowels (/a/, /e/, /u/), six sentences, a phonetically balanced text, and spontaneous speech. This study uses only the sustained vowel /a/, with each repetition treated as a separate sample, yielding 1,498 audio signals.  

\item Toronto Emotional Speech Set (TESS) \cite{pichora2020toronto} comprises 2,800 utterances from two female speakers (aged 26 and 64). Each produced 200 target words across seven emotions: anger, disgust, fear, happiness, pleasant surprise, sadness, and neutral. Recordings were made at 24~kHz and are widely used in speech emotion recognition research.  
\end{itemize}

Audio from both datasets was recorded in controlled environments with minimal background noise and is therefore considered clean; throughout this paper, we refer to these uncorrupted recordings as ‘clean audio'. Controlled corruptions are later introduced to simulate specific noise conditions, allowing a systematic investigation of QNN robustness under well-defined corruptions rather than using available noisy datasets. The use of noisy datasets will be explored in future work. All audio signals are subsequently converted into log-Mel spectrograms, as described in Section~\ref{subsection: speech process}, and used as inputs to the classifier models.

\subsection{Four Types of Corruption}
We evaluate model robustness under four common acoustic corruptions: Gaussian noise, pitch shift, temporal shift, and speed variation. Each corruption is applied at six severity levels. For Gaussian noise, pitch shift, and temporal shift, $\sigma = 0$ denotes no corruption, while for speed variation $\sigma = 1$ corresponds to the unmodified signal (see Table~\ref{tab:scaling_factor}). \noindent
To enable a fair and quantifiable robustness evaluation, we define controlled corruption severity levels for each perturbation type, as detailed in Table~\ref{tab:scaling_factor}. Each severity parameter was chosen to produce progressively stronger yet physically realistic degradations of the input waveform, with ranges selected to cover typical variations observed in natural speech.

\begin{table}[]
\caption{Scaling factor of each corruption type}
\label{tab:scaling_factor}
\resizebox{\columnwidth}{!}{%
\begin{tabular}{@{}lll@{}}
\toprule
Corruption Type & Clean value      & Severity level                                        \\ \midrule
Gaussian noise  & \(\sigma_n\) = 0 & \(\sigma_n\) = {[}0.01, 0.05, 0.1, 0.15, 0.2, 0.25{]} \\
Pitch shift     & \(\sigma_p\) = 0 & \(\sigma_p\) = {[}0.05, 0.1, 0.15, 0.2, 0.25, 0.3{]}   \\
Temporal shift & \(\sigma_t\) = 0 & \(\sigma_t\) = {[}0.025, 0.05, 0.075, 0.1, 0.125, 0.15{]} \\
Speed variation & \(\sigma_s\) = 1 & \(\sigma_s\) = {[}1.05, 1.1, 1.15, 1.2, 1.25, 1.3{]}  \\ \bottomrule
\end{tabular}%
}
\end{table}

\textbf{Gaussian noise:} To simulate background interference, zero-mean Gaussian noise with standard deviation $\sigma_n$ is applied directly to the waveform, as described in Algorithm \ref{alg: gaussian noise}. The standard deviation $\sigma_{n}$ is normalized with respect to the signal amplitude, ranging from $0.01$ to $0.25$, thereby simulating signal-to-noise ratios (SNR) between roughly $40$ dB and $12$ dB values representative of mild to moderate environmental noise. This corruption perturbs each sample value independently, thereby degrading the signal without altering its global structure.  

\begin{algorithm}[H]
\caption{Gaussian Noise}
\label{alg: gaussian noise}
\begin{algorithmic}[1]
\INPUT Audio signals $\mathcal{A}=\{a^{(i)}\}_{i=1}^{N}$, where $a^{(i)} \in \mathbb{R}^{L_i}$, and a noise factor $\sigma_n$
\OUTPUT Noisy signals $\tilde{\mathcal{A}}=\{\tilde{a}^{(i)}\}_{i=1}^{N}$, with $|\tilde{a}^{(i)}| = L_i$
\For{each $a^{(i)} \in \mathcal{A}$}
  \State Compute standard deviation $\hat{\sigma}_i \gets \mathrm{std}(a^{(i)})$
  \State Sample Gaussian noise: $z^{(i)} = [z^{(i)}_1,\ldots,z^{(i)}_{L_i}]$ with $z^{(i)}_k \sim \mathcal{N}(0,\sigma_n)$ independently
  \State Add scaled noise: $\tilde{a}^{(i)} \gets a^{(i)} + \hat{\sigma}_i \cdot z^{(i)}$
  \State Clamp amplitudes: $\tilde{a}^{(i)} \gets \mathrm{clip}(\tilde{a}^{(i)}, -1, 1)$
\EndFor
\State \Return $\tilde{\mathcal{A}}$
\end{algorithmic}
\end{algorithm}

\begin{algorithm}[H]
\caption{Pitch Shift}
\label{alg: pitch shift}
\begin{algorithmic}[1]
\INPUT Audio signals $\mathcal{A}=\{a^{(i)}\}_{i=1}^{N}$ with $a^{(i)}\in\mathbb{R}^{L_i}$; sample rate $\mathrm{sr}$; semitone factor $\sigma_p$
\OUTPUT Pitch-shifted signals $\tilde{\mathcal{A}}=\{\tilde{a}^{(i)}\}_{i=1}^{N}$
\For{each $a^{(i)} \in \mathcal{A}$}
  \State Sample semitone shift $\Delta_i \sim \mathcal{N}(0,\sigma_p)$
  \State $\tilde{a}^{(i)} \gets \mathrm{PitchShift}\!\big(a^{(i)},\, \Delta_i,\, \mathrm{sr}\big)$ \Comment{using \texttt{librosa.effects.pitch\_shift})}
\EndFor
\State \Return $\tilde{\mathcal{A}}$
\end{algorithmic}
\end{algorithm}
\begin{algorithm}[H]
\caption{Temporal Shift}
\label{alg: temporal shift}
\begin{algorithmic}[1]
\INPUT Audio signals $\mathcal{A}=\{a^{(i)}\}_{i=1}^{N}$ with $a^{(i)}\in\mathbb{R}^{L_i}$; maximum shift proportion $\sigma_t$
\OUTPUT Shifted signals $\tilde{\mathcal{A}}=\{\tilde{a}^{(i)}\}_{i=1}^{N}$ with $|\tilde{a}^{(i)}|=L_i$
\For{each $a^{(i)} \in \mathcal{A}$}
  \State Sample integer shift $s_i \sim \mathcal{N}(0,\sigma_t)$
  \If{$s_i>0$} \Comment{right shift: leading zeros, drop tail}
    \State $\tilde{a}^{(i)} \gets [\underbrace{0,\ldots,0}_{s_i},\, a^{(i)}_{1{:}L_i-s_i}]$
  \ElsIf{$s_i<0$} \Comment{left shift: trailing zeros, drop head}
    \State $\tilde{a}^{(i)} \gets [a^{(i)}_{1{-}s_i{:}L_i},\, \underbrace{0,\ldots,0}_{-s_i}]$
  \Else
    \State $\tilde{a}^{(i)} \gets a^{(i)}$
  \EndIf
\EndFor
\State \Return $\tilde{\mathcal{A}}$
\end{algorithmic}
\end{algorithm}
\textbf{Pitch shift.} To mimic variations in speaker pitch or tonal shifts, the fundamental frequency of the audio is modified by $\sigma_p$ pitch factor while preserving temporal alignment (Algorithm \ref{alg: pitch shift}). On the spectrogram, this manifests as a vertical translation of harmonic energy across frequency bins. The pitch factor $\sigma_{p} \in [0.05,\,0.3]$ produce frequency scaling factors within $\pm3\%$ (0.5 semitone).

\textbf{Temporal shift:} To simulate timing misalignments, the waveform is randomly shifted along the time axis by a displacement proportion $\sigma_t$ of its total length (Algorithm~\ref{alg: temporal shift}). A positive $\sigma_t$ delays the signal (right shift), whereas a negative $\sigma_t$ advances it (left shift). Zero-padding is applied to maintain the original signal length. On the spectrogram, this corresponds to a horizontal translation of consecutive time frames. Severity $\sigma_{t} \in [0.025,\,0.15]$ defines the maximum displacement, ensuring that the waveform is shifted by at most $15\%$ of its duration.

\textbf{Speed variation:} To emulate speaking rate variability, the signal is time-stretched by a factor $r = e^u$ where $u \sim \mathcal{N}(0, \log(\sigma_s))$ (Algorithm \ref{alg: speed variation}). This log-normal distribution ensures symmetric treatment of acceleration and deceleration around the original playback rate. With $\sigma_{s} \in [1.05,\,1.3]$, the logarithmic standard deviation ranges from approximately $0.049$ to $0.262$, producing speed modifications that typically fall within $\pm5$--$30\%$.

\begin{algorithm}[H]
\caption{Speed Variation}
\label{alg: speed variation}
\begin{algorithmic}[1]
\INPUT Audio signals $\mathcal{A}=\{a^{(i)}\}_{i=1}^{N}$ with $a^{(i)} \in \mathbb{R}^{L_i}$, speed factor $\sigma_s$
\OUTPUT Speed-changed signals $\tilde{\mathcal{A}}=\{\tilde{a}^{(i)}\}_{i=1}^{N}$ with $|\tilde{a}^{(i)}| = L_i$
\For{each $a^{(i)} \in \mathcal{A}$}
  \State Sample log-speed: $u \sim \mathcal{N}(0,\log(\sigma_s))$
  \State Convert to speed ratio: $r \gets e^u$
  \State Apply time-stretch: $\hat{a}^{(i)} \gets \mathrm{TimeStretch}(a^{(i)}, r)$
  \If{$|\hat{a}^{(i)}| > L_i$}
     \State Truncate: $\hat{a}^{(i)} \gets \hat{a}^{(i)}[1{:}L_i]$
  \ElsIf{$|\hat{a}^{(i)}| < L_i$}
     \State Pad with zeros: $\hat{a}^{(i)} \gets \mathrm{Pad}(\hat{a}^{(i)}, L_i)$
  \EndIf
  \State Append: $\tilde{a}^{(i)} \gets \hat{a}^{(i)}$
\EndFor
\State \Return $\tilde{\mathcal{A}}$
\end{algorithmic}
\end{algorithm}
\vspace{-1em}
\subsection{Training and Testing procedures}
Each dataset was partitioned 65\%/15\%/20\% into training/validation/testing. Both QNN and CNN models were trained on the same dataset. To assess robustness in a clean-train/corrupted-test setting, each corruption was applied only to the held-out test set at evaluation time. Model performance was further validated through 10 independent runs (seeds) with different training–testing splits and randomized parameter initializations such as corruption random, and weight initializations to support generalization and reliable assessment. This protocol (Fig. \ref{fig:framework}) provides a comprehensive evaluation of model performance across corruption types and severities. Each model was trained for up to 10,000 epochs with a batch size of 20, using Adam (learning rate at $10^{-5}$, weight decay at $10^{-2}$). Early stopping was applied if validation loss did not improve for 30 consecutive epochs, and the best-validation checkpoint was used for testing. All quantum layers were simulated on a classical device using PennyLane \cite{bergholm2018pennylane}. The quanvolution implementation uses JAX for just-in-time compilation and vectorization to reduce compilation time \cite{jax2018github}. Model implementation was carried out in PyTorch 2.7.0 with CUDA 12.6 support and PennyLane 0.38.0. Training was conducted on a computer with an AMD Ryzen 5 9600X 6-core CPU and an NVIDIA RTX 4060 Ti GPU with 8 GB of GDDR6 VRAM.
\subsection{Evaluation Metrics}

To assess model performance, we first compute classification accuracy on the clean test set and across each severity level of the four corruption types. This allows us to observe how accuracy changes between clean and corrupted conditions, as visualized in Section ~\ref{subsec:Robustness Analysis}.  

To provide a standardized and comprehensive robustness evaluation under input degradations, we adopt the Corruption Error (CE), mean Corruption Error (mCE), Relative Corruption Error (RCE), and Relative mean Corruption Error (RmCE) metrics, following \cite{hendrycks2019benchmarking}. These metrics enable a fair comparison of robustness between QNNs and CNNs, as reported in Section \ref{subsec:Robustness Analysis}.  
The CE metric quantifies robustness by normalizing the error rate of a given model ($E_{i,s}$) with respect to the error rate of a baseline model ($E^{\text{base}}_{i,s}$) across all severity levels of corruption type $i$, as defined in Eq.~\ref{eq:ce}:

\begin{equation}
    \mathrm{CE}_{i} =
    \frac{\sum_{s=1}^{6} \mathrm{E}_{i,s}}
         {\sum_{s=1}^{6} \mathrm{E}^{\text{base}}_{i,s}}
    =
    \frac{\sum_{s=1}^{6} \left( 1 - \mathrm{Acc}_{i,s} \right)}
         {\sum_{s=1}^{6} \left( 1 - \mathrm{Acc}^{\text{base}}_{i,s} \right)},
    \label{eq:ce}
\end{equation}
here, $\mathrm{Acc}_{i,s}$ denotes the accuracy of model $i$ at severity level $s \in [1,6]$, while $\mathrm{Acc}^{\text{base}}_{i,s}$ represents the accuracy of the baseline model. To measure robustness across all $N=4$ corruption types, we compute the mean Corruption Error (mCE), as shown in Eq.~\ref{eq:mce}:

\begin{equation}
    \mathrm{mCE} = \frac{1}{N} \sum_{i=1}^{N} \mathrm{CE}_{i}.
    \label{eq:mce}
\end{equation}

While CE and mCE capture baseline-normalized performance under corruption, they do not reflect how much a model degrades relative to its own clean accuracy. To evaluate this aspect, we use the Relative Corruption Error (RCE), which measures robustness degradation compared to the clean baseline, as defined in Eq.~\ref{eq:rce}:

\begin{equation}
    \mathrm{RCE}_{i} =
    \frac{\sum_{s=1}^{6} 
    \left(\mathrm{Acc}_{\text{clean}} - \mathrm{Acc}_{i,s}\right)}
         {\sum_{s=1}^{6} 
    \left(\mathrm{Acc}^{\text{base}}_{\text{clean}} - \mathrm{Acc}^{\text{base}}_{i,s}\right)},
    \label{eq:rce}
\end{equation}
here, $\mathrm{Acc}_{\text{clean}}$ is the model accuracy on the clean test set. Extending this to all corruption types, the Relative mean Corruption Error (RmCE) provides an overall measure of relative robustness, as expressed in Eq.~\ref{eq:rmce}:

\begin{equation}
    \mathrm{RmCE} = \frac{1}{N} \sum_{i=1}^{N} \mathrm{RCE}_{i}.
    \label{eq:rmce}
\end{equation}

In this work, CNN-Base serves as the baseline model. Therefore, its CE, mCE, RCE, and RmCE values are normalized to 1. A QNN exhibiting values lower than 1 indicates greater robustness than CNN-Base, whereas values greater than 1 indicate reduced robustness.  

Additionally, we employ the confusion matrix to analyse classification outcomes across emotions (Section.~\ref{subsec: confusion matrix}), enabling identification of which emotions are most degraded under corruption and which remain most reliably classified. Furthermore, the training dynamics, including training loss and validation accuracy per epoch, are examined to assess model convergence speed (Section.~\ref{subsec: convergence}). The influence of quantum circuit depth on QNN performance is also investigated in the ablation study (Section~\ref{subsec: ablation study}). Results in other sections present and discuss the best-performing configurations identified in this study (see Table \ref{tab:depth_circuit}). 
\section{Results and Discussion}
\label{sec: results and discussion}
\subsection{Robustness Analysis}
\label{subsec:Robustness Analysis}
\begin{figure*}[!t]
    \centering
    \begin{subfigure}[b]{0.9\textwidth}
        \centering
        \includegraphics[width=\textwidth]{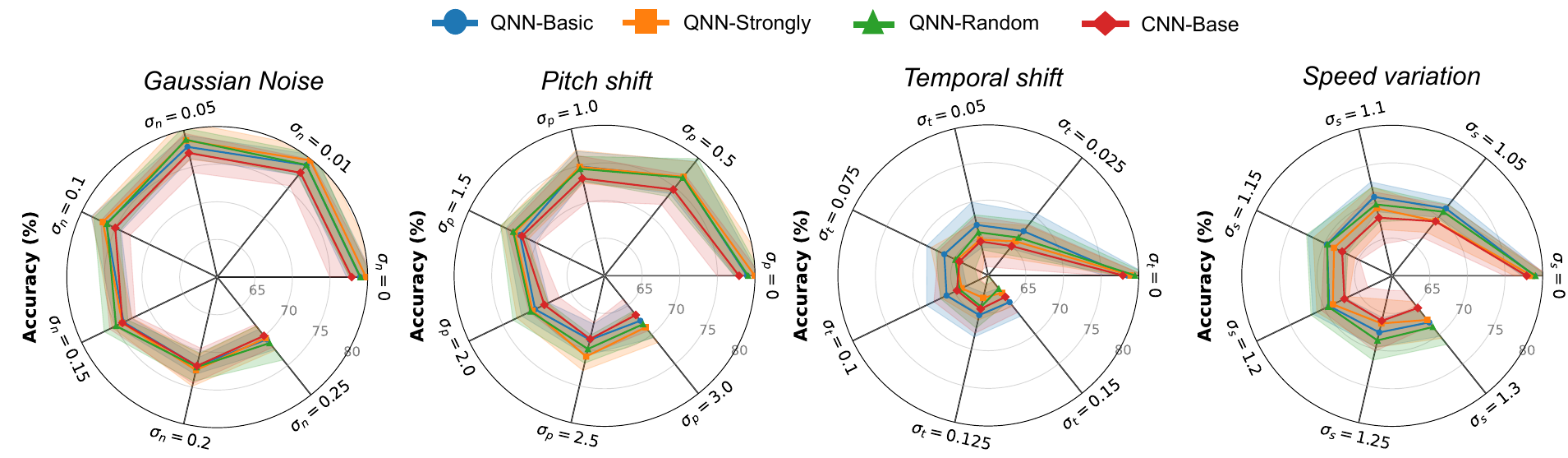}
        \caption{AVFAD Dataset}
        \label{fig: avfad_radar}
    \end{subfigure}
    \begin{subfigure}[b]{0.9\textwidth}
        \centering
        \includegraphics[width=\textwidth]{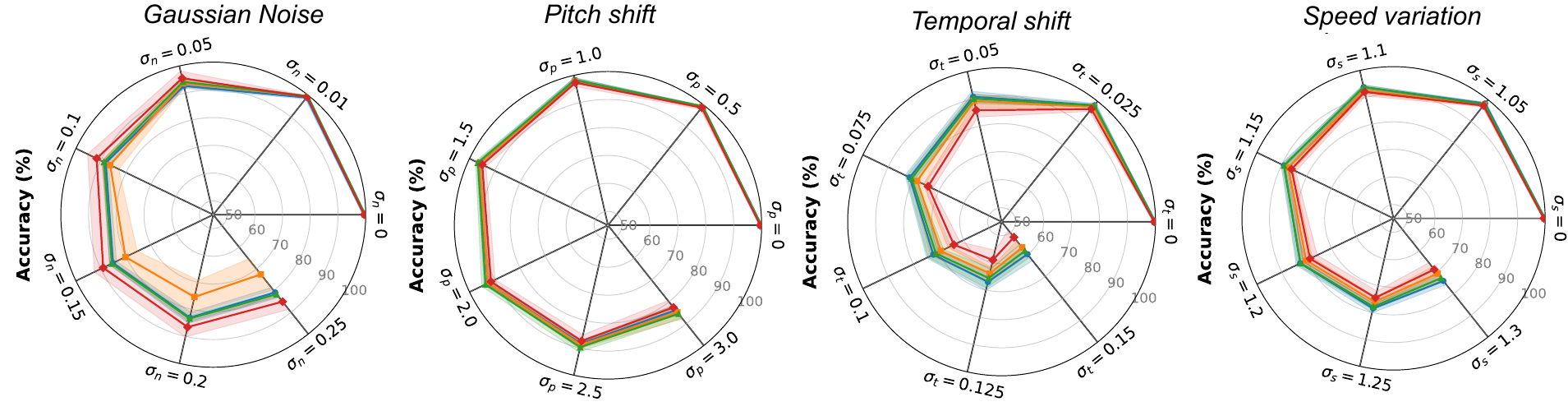}
        \caption{TESS Dataset}
        \label{fig: TESS_radar}
    \end{subfigure}
    \caption{Accuracy of QNNs (QNN-Basic, QNN-Strongly, and QNN-Random) and CNN-Base under four corruption types: Gaussian noise, pitch shift, temporal shift, and speed variation on AVFAD, TESS datasets. The corresponding circuit depths ($d$) for QNN-Basic, QNN-Strongly, and QNN-Random are summarized in Table~\ref{tab:depth_circuit}.}
    \label{fig: radar}
    \vspace{-10pt}
\end{figure*}
\begin{table*}[!t]
\centering
\caption{Comparison robustness metrics of CNN-Base and QNN Models on AVFAD and TESS datasets, with values reported as the mean $\pm$ standard deviation across 10 random seeds. The corresponding circuit depths ($d$) for QNN-Basic, QNN-Strongly, and QNN-Random are summarized in Table~\ref{tab:depth_circuit}.}
\label{tab:cnn_qnn}
\resizebox{0.85\textwidth}{!}{%
\begin{tabular}{@{}lcccccccc@{}}
\toprule
\multicolumn{9}{c}{AVFAD dataset} \\ \midrule
\multicolumn{1}{c|}{\multirow{2}{*}{Corruption Type}} &
  \multicolumn{4}{c|}{CE} &
  \multicolumn{4}{c}{RCE} \\ \cmidrule(l){2-9} 
\multicolumn{1}{c|}{} &
  CNN-Base &
  QNN-Basic &
  QNN-Random &
  \multicolumn{1}{c|}{QNN-Strongly} &
  CNN-Base &
  QNN-Basic &
  QNN-Random &
  QNN-Strongly \\ \midrule
\multicolumn{1}{l|}{Gaussian noise} &
  1.00 ± 0.00 &
  0.98 ± 0.08 &
  \textbf{0.96 ± 0.07} &
  \multicolumn{1}{c|}{0.96 ± 0.10} &
  1.00 ± 0.00 &
  1.69 ± 1.68 &
  1.86 ± 2.64 &
  \textbf{1.68 ± 1.44} \\
\multicolumn{1}{l|}{Pitch shift} &
  1.00 ± 0.00 &
  0.96 ± 0.06 &
  0.95 ± 0.07 &
  \multicolumn{1}{c|}{\textbf{0.94 ± 0.06}} &
  1.00 ± 0.00 &
  1.08 ± 0.46 &
  \textbf{0.99 ± 0.27} &
  1.08 ± 0.30 \\
\multicolumn{1}{l|}{Temporal shift} &
  1.00 ± 0.00 &
  \textbf{0.95 ± 0.04} &
  1.00 ± 0.08 &
  \multicolumn{1}{c|}{1.01 ± 0.08} &
  1.00 ± 0.00 &
  \textbf{0.93 ± 0.16} &
  1.19 ± 0.41 &
  1.19 ± 0.44 \\
\multicolumn{1}{l|}{Speed variation} &
  1.00 ± 0.00 &
  \textbf{0.93 ± 0.04} &
  0.93 ± 0.08 &
  \multicolumn{1}{c|}{0.97 ± 0.05} &
  1.00 ± 0.00 &
  0.95 ± 0.39 &
  \textbf{0.95 ± 0.30} &
  1.02 ± 0.36 \\ \midrule
\multicolumn{1}{l|}{mCE} &
  1.00 ± 0.00 &
  \textbf{0.96 ± 0.03} &
  0.96 ± 0.04 &
  \multicolumn{1}{c|}{0.97 ± 0.04} &
  - &
  - &
  - &
  - \\
\multicolumn{1}{l|}{RmCE} &
  - &
  - &
  - &
  \multicolumn{1}{c|}{-} &
  1.00 ± 0.00 &
  \textbf{1.16 ± 0.62} &
  1.25 ± 0.79 &
  1.24 ± 0.53 \\ \midrule
\multicolumn{9}{c}{TESS dataset} \\ \midrule
\multicolumn{1}{l|}{\multirow{2}{*}{Corruption Type}} &
  \multicolumn{4}{c|}{CE} &
  \multicolumn{4}{c}{RCE} \\ \cmidrule(l){2-9} 
\multicolumn{1}{l|}{} &
  CNN-Base &
  QNN-Basic &
  QNN-Random &
  \multicolumn{1}{c|}{QNN-Strongly} &
  CNN-Base &
  QNN-Basic &
  QNN-Random &
  QNN-Strongly \\ \midrule
\multicolumn{1}{l|}{Gaussian noise} &
  1.00 ± 0.00 &
  1.52 ± 0.55 &
  \textbf{1.40 ± 0.42} &
  \multicolumn{1}{c|}{1.99 ± 0.93} &
  1.00 ± 0.00 &
  1.59 ± 0.67 &
  \textbf{1.51 ± 0.53} &
  2.16 ± 1.14 \\
\multicolumn{1}{l|}{Pitch shift} &
  1.00 ± 0.00 &
  0.89 ± 0.16 &
  \textbf{0.79 ± 0.13} &
  \multicolumn{1}{c|}{0.88 ± 0.13} &
  1.00 ± 0.00 &
  0.94 ± 0.14 &
  \textbf{0.83 ± 0.13} &
  0.90 ± 0.13 \\
\multicolumn{1}{l|}{Temporal shift} &
  1.00 ± 0.00 &
  \textbf{0.78 ± 0.05} &
  0.81 ± 0.02 &
  \multicolumn{1}{c|}{0.86 ± 0.10} &
  1.00 ± 0.00 &
  \textbf{0.79 ± 0.05} &
  0.82 ± 0.02 &
  0.87 ± 0.10 \\
\multicolumn{1}{l|}{Speed variation} &
  1.00 ± 0.00 &
  \textbf{0.82 ± 0.07} &
  0.84 ± 0.07 &
  \multicolumn{1}{c|}{0.94 ± 0.13} &
  1.00 ± 0.00 &
  \textbf{0.84 ± 0.07} &
  0.86 ± 0.06 &
  0.96 ± 0.12 \\ \midrule
\multicolumn{1}{l|}{mCE} &
  1.00 ± 0.00 &
  1.00 ± 0.11 &
  \textbf{0.96 ± 0.08} &
  \multicolumn{1}{c|}{1.17 ± 0.19} &
  - &
  - &
  - &
  - \\
\multicolumn{1}{l|}{RmCE} &
  - &
  - &
  - &
  \multicolumn{1}{c|}{-} &
  1.00 ± 0.00 &
  1.04 ± 0.14 &
  \textbf{1.01 ± 0.11} &
  1.22 ± 0.24 \\ \bottomrule
\end{tabular}%
}

\end{table*}
Fig. \ref{fig: radar} shows the accuracies of three QNNs and CNN-Base across four corruption types, while Table \ref{tab:cnn_qnn} presents robustness metrics such as CE, mCE, RCE, RmCE of QNNs and CNN-Base of two datasets. As expected, the accuracies of all models gradually degrade as corruption severity increases. 
\subsubsection{Gaussian noise}
On the AVFAD data, QNNs maintain slightly higher accuracy than CNN-Base for most level of noise, with all three QNN models performing comparably. Their smaller CE values further confirm improved corruption robustness compared to CNN-Base. However, at the most severe levels, CNN-Base has the same accuracy as QNNs. Because CNN-Base has a lower clean accuracy, its relative degradation is less pronounced, leading to smaller RCE values compared to QNNs. Thus, QNNs demonstrate superior corruption robustness but inferior relative robustness on AVFAD. 

In contrast, on the TESS dataset, CNN-Base clearly outperforms all QNNs under Gaussian noise. Notably, the accuracy of QNN-Strongly significantly drops from 100\% to 70\%, corresponding to CE = 1.99, and RCE = 2.16, whereas CNN-Base attains over 85\% accuracy across all noise levels. These results highlight that QNNs are less resilient to Gaussian noise in emotional speech classification. This vulnerability can be explained that the noise could be amplified random fluctuation at the quantum encoding of the quanvolutional layer. Additionally, the performance differences between the two datasets arise because AVFAD voice recordings are more irregular and noisy than TESS, making QNNs relatively competitive with CNN-Base on this challenging data. In contrast, TESS speech is cleaner and more structured. Consequently, additive Gaussian noise disrupts the quanvolutional layers more severely, giving CNN-Base a larger performance advantage.

\subsubsection{Pitch shift}
With increasing pitch shift severity, QNNs consistently achieve higher accuracy and improved robustness compared to CNN-Base across both datasets, as indicated by their smaller CE values. This advantage is more pronounced on the TESS dataset than on AVFAD. For example, QNN-Random achieves CE = 0.79 and RCE = 0.83 on TESS, compared to CE = 0.95 and RCE = 0.99 on AVFAD. These findings suggest that QNNs are more effective at capturing relational and structural frequency patterns, thereby enhancing their resilience to spectral-domain corruptions such as pitch shift. 

\subsubsection{Temporal shift}
Among four corruptions types, temporal shifting severely disrupt model accuracy in the most degradation, with accuracies dropping by 20\%-50\%. On the AVFAD dataset, QNN-Basic obtains the highest accuracy, particularly at moderate cropping levels ($\sigma_t = 0.075$–$0.125$), with smaller CE (0.95) and RCE (0.93) values compared to CNN-Base. By contrast, QNN-Random and QNN-Strongly perform similarly to CNN-Base in terms of CE but show worse relative robustness with RCE = 1.19.

On the TESS dataset, QNNs consistently outperform CNN-Base with increasing cropping severity, as demonstrated by their smaller CE values (0.78-0.86), RCE values (0.79-0.87). At the most severe level, QNNs achieve approximately 8\% accuracy than CNN-Base. QNN-Basic once again achieves the strongest performance, with the lowest CE (0.78) and RCE (0.79), corresponding to a 22\% robustness improvement over CNN-Base. 
These findings suggest that QNNs are more effective at preserving global temporal features, while CNN-Base degrades sharply when temporal continuity is disrupted. 

\subsubsection{Speed variation}
Speed variation further highlights the robustness of QNNs compared to CNN-Base across both datasets. On the AVFAD dataset, QNN-Basic and QNN-Random achieve the highest accuracies, with smaller CE and RCE values than CNN-Base, indicating improved robustness. On the TESS dataset, QNN-Basic demonstrates the strongest performance, consistently attaining the lowest CE and RCE values among all models. These results can demonstrate that quantum mappings are less sensitive to rescaling of time features. 

Overall, QNNs demonstrate consistently improved robustness over CNN-Base for pitch shift, temporal shifting, and speed variation, while CNN-Base remains more resilient under Gaussian noise. The robustness metrics further reveal that different quantum models respond differently to corruption types. When evaluated across all four corruptions using mCE and RmCE, QNN-Basic achieves the best overall robustness on the AVFAD dataset, whereas QNN-Random performs best on the TESS dataset. In contrast, QNN-Strongly, with its more complex entanglement structure, exhibits reduced resilience compared to the other quantum models. The fact that the RmCE values of QNNs exceed CNN-Base (RmCE $>$ 1) is mainly due to their poor relative robustness under Gaussian noise, which dominates the mean value.

\begin{figure*}[!t]
    \centering
    \includegraphics[width=\textwidth]{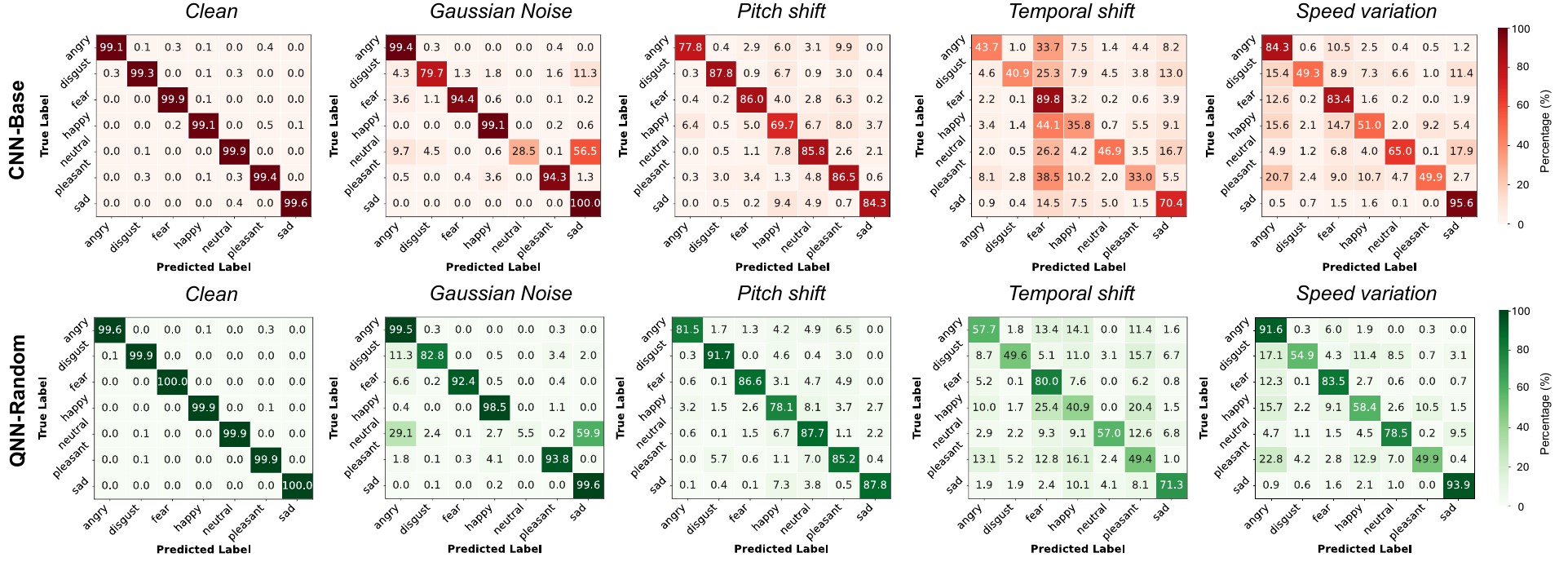}
    \caption{Confusion matrix for CNN-Base (red) and QNN-Random (green) on the clean and under the most severe level of four corruption types. The corresponding circuit depth ($d$) of QNN-Random is $d=1$.}
    \vspace{-10pt}
    \label{fig:tess_qnn_cnn_confusion_matrix}
\end{figure*}
\subsection{Robustness Comparison with Large-Scale Classical Networks}
\label{subsec:QNN_vs_Advanced}
Table~\ref{tab:mce_rmce_complex_model} compares the robustness of QNNs with advanced classical architectures, ResNet-18 and VGG-16, across the AVFAD and TESS datasets using the mCE and RmCE metrics. Despite having roughly 1M parameters (+4 qubits) compared with 11M and 134M for ResNet-18 and VGG-16, respectively, all QNN models achieve comparable or superior robustness under range of input corruptions. On the AVFAD dataset, QNN models maintain mCE = 0.96 , close to or below the large classical networks, while avoiding the high RmCE variance observed in VGG-16 ($1.95 \pm 1.29$). On the TESS dataset, the advantage of quantum encoding becomes more evident: deeper classical networks degrade markedly (ResNet-18: mCE = $1.17 \pm 0.25$ , VGG-16: mCE = $1.23 \pm 0.39$), whereas QNN-Random achieves the lowest mCE ($0.96 \pm 0.08$) and RmCE ($1.01 \pm 0.11$). These results indicate that quantum feature mappings yield more invariant and compact representations, enabling QNNs to reach or exceed the robustness of much larger classical architectures while maintaining a higher parameter efficiency..
\begin{table}[]
\centering
\caption{Comparison of model robustness (mCE and RmCE) between classical architectures (CNN-Base, ResNet-18, VGG-16) and QNNs on the AVFAD and TESS datasets. The corresponding circuit depths ($d$) for QNN-Basic, QNN-Strongly, and QNN-Random are summarized in Table~\ref{tab:depth_circuit}.}
\label{tab:mce_rmce_complex_model}
\resizebox{\columnwidth}{!}{%
\begin{tabular}{@{}cccccc@{}}
\toprule
\multirow{2}{*}{Model\_Name} & \multirow{2}{*}{Model Size} & \multicolumn{2}{c}{AVFAD dataset} & \multicolumn{2}{c}{TESS dataset} \\ \cmidrule(l){3-6} 
              &                  & mCE         & RmCE        & mCE         & RmCE        \\ \midrule
CNN-Base           & 1M               & 1.00 ± 0.00 & 1.00 ± 0.00 & 1.00 ± 0.00 & 1.00 ± 0.00 \\
ResNet-18     & 11M              & 1.06 ± 0.04 & 1.82 ± 1.59 & 1.17 ± 0.25 & 1.25 ± 0.28 \\
VGG-16        & 134M             & 0.94 ± 0.09 & 1.95 ± 1.29 & 1.23 ± 0.39 & 1.28 ± 0.35 \\ \midrule
QNN-Basic    & 1M + 4q (frozen) & 0.96 ± 0.03 & 1.16 ± 0.62 & 1.00 ± 0.11 & 1.04 ± 0.14 \\
QNN-Random   & 1M + 4q (frozen) & 0.96 ± 0.04 & 1.25 ± 0.79 & 0.96 ± 0.08 & 1.01 ± 0.11 \\
QNN-Strongly & 1M + 4q (frozen) & 0.97 ± 0.04 & 1.24 ± 0.53 & 1.17 ± 0.19 & 1.22 ± 0.24 \\ \bottomrule
\end{tabular}%
}
\end{table}

\subsection{Multiple classes accuracy}
\label{subsec: confusion matrix}
The confusion matrices in Fig.~\ref{fig:tess_qnn_cnn_confusion_matrix} illustrate class-wise accuracies on the TESS dataset, evaluated on both clean testing data and under the most severe level of each corruption. For clarity, we focus on CNN-Base and QNN-Random, which is taken as representative of the QNN models; the corresponding confusion matrices for QNN-Basic and QNN-Strongly are provided in the Supplementary Material. Across the seven emotions, both QNN and CNN-Base achieve nearly perfect accuracy on clean data, with QNN showing marginally stronger generalization. However, performance degrades substantially under corrupted conditions.
Under Gaussian noise, the \textit{neutral} emotion is most severely affected, with accuracy dropping from 99.9\% to 28.5\% for CNN-Base and from 99.9\% to 5.5\% for QNN, as both models frequently misclassify \textit{neutral} as \textit{sad}. This occurs because Gaussian noise strongly overlaps with the flat spectral patterns of neutral speech, making it difficult for the models to distinguish \textit{neutral} from \textit{sad}, which has similar low-energy frequency distributions. In contrast, both models classify \textit{sad}, \textit{happy}, and \textit{angry} reliably. 
For pitch shift, the \textit{happy} emotion is most affected, while \textit{disgust} remains the most robust class. This is because happy speech relies heavily on pitch elevation as a discriminative feature, making it highly sensitive to spectral shifts \cite{yuan2002acoustic}. However, QNN still classifies \textit{happy} with relative high accuracy (78.1\%), compared to 69.7\% for CNN-Base. 
With temporal shifting, most emotions experience sharp degradation except for \textit{fear}, which reaches 89.8\% in CNN-Base and 80\% in QNN. The robustness of \textit{fear} arises because it is associated with sustained high-energy segments, so even partial signals retain sufficient cues for classification. The high accuracy is also influenced by a prediction bias, which both models tend to misclassifying cropped samples as \textit{fear}, and this bias is stronger in CNN-Base than in QNN.
For speed variation, the most affected emotions are \textit{disgust}, \textit{happy}, and \textit{pleasant}, while \textit{angry}, \textit{fear}, and \textit{sad} remain classified with relatively high accuracy. This is because speed corruption disproportionately distort timing-dependent cues, such as rhythmic variations and prosodic changes, which are essential for emotions like \textit{happy} and \textit{disgust} \cite{rigoulot2013feeling}. By contrast, \textit{angry} and \textit{fear} rely on more global energy and spectral patterns, which are less sensitive to time-scale distortions.

In general, \textit{fear} emerges as the most robustly classified emotion across corruptions, while \textit{neutral} is most vulnerable under Gaussian noise, and \textit{happy} is most vulnerable under pitch shift, temporal shifting, and speed variation. Among the models, QNN-Random consistently outperforms CNN-Base across most corruptions, with the notable exception of the \textit{neutral} class under Gaussian noise, where CNN-Base exhibits higher robustness.

\subsection{Ablation Study}
\label{subsec: ablation study}
To evaluate the impact of quantum circuit depth ($d$) on the robustness of QNNs, we conducted an ablation study by varying $d$ from 1 to 50. Table \ref{tab:depth_circuit} summarizes the optimal circuit depth for the three quantum circuit architecture: BEQC in QNN-Basic ($d_{\mathrm{BEQC}}$), SEQC in QNN-Strongly ($d_{\mathrm{SEQC}}$), and RQC in QNN-Random ($d_{\mathrm{RQC}}$) across four corruption types on the AVFAD and TESS datasets. As shown, there is no consistent trend in the optimal depth across all settings. QNN-Random consistently performs best at the shallow depth of $d_{\mathrm{RQC}} = 1$, regardless of dataset or corruption type, indicating strong robustness even with minimal circuit complexity. In contrast, QNN-Basic and QNN-Strongly exhibit variable optimal depths depending on the corruption type and dataset. For instance, under Gaussian noise, QNN-Basic favours deeper circuits ($d_{\mathrm{BEQC}} = 30$–50), whereas QNN-Strongly achieves optimal performance at moderate depths ($d_{\mathrm{SEQC}} = 10$–20).
\begin{table}[]
\centering
\caption{The optimal circuit depths for the three quantum circuit architectures: BEQC in QNN-Basic ($d_{\mathrm{BEQC}}$), SEQC in QNN-Strongly ($d_{\mathrm{SEQC}}$), and RQC in QNN-Random ($d_{\mathrm{RQC}}$).}
\label{tab:depth_circuit}
\resizebox{0.8\columnwidth}{!}{%
\begin{tabular}{@{}cccc@{}}
\toprule
\multicolumn{4}{c}{AVFAD dataset} \\ \midrule
Corruption Type &
  \begin{tabular}[c]{@{}c@{}}QNN-Basic\\ $d_{BEQC}$\end{tabular} &
  \begin{tabular}[c]{@{}c@{}}QNN-Strongly\\ $d_{SEQC}$\end{tabular} &
  \begin{tabular}[c]{@{}c@{}}QNN-Random\\ $d_{RQC}$\end{tabular} \\ \midrule
Gaussian noise &
  30 &
  20 &
  25 \\
Pitch shift &
  4 &
  15 &
  15 \\
Temporal shift &
  15 &
  10 &
  1 \\
Speed variation &
  4 &
  1 &
  25 \\ \midrule
\multicolumn{4}{c}{TESS dataset} \\ \midrule
Corruption Type &
  \begin{tabular}[c]{@{}c@{}}QNN-Basic\\ $d_{BEQC}$\end{tabular} &
  \begin{tabular}[c]{@{}c@{}}QNN-Strongly\\ $d_{SEQC}$\end{tabular} &
  \begin{tabular}[c]{@{}c@{}}QNN-Random\\ $d_{RQC}$\end{tabular} \\ \midrule
Gaussian noise &
  50 &
  10 &
  1 \\
Pitch shift &
  30 &
  30 &
  1 \\
Temporal shift &
  15 &
  50 &
  1 \\
Speed variation &
  15 &
  4 &
  1 \\ \bottomrule
\end{tabular}%
}
\end{table}
\begin{figure}[htbp]
    \centering
    \includegraphics[width=0.45\textwidth]{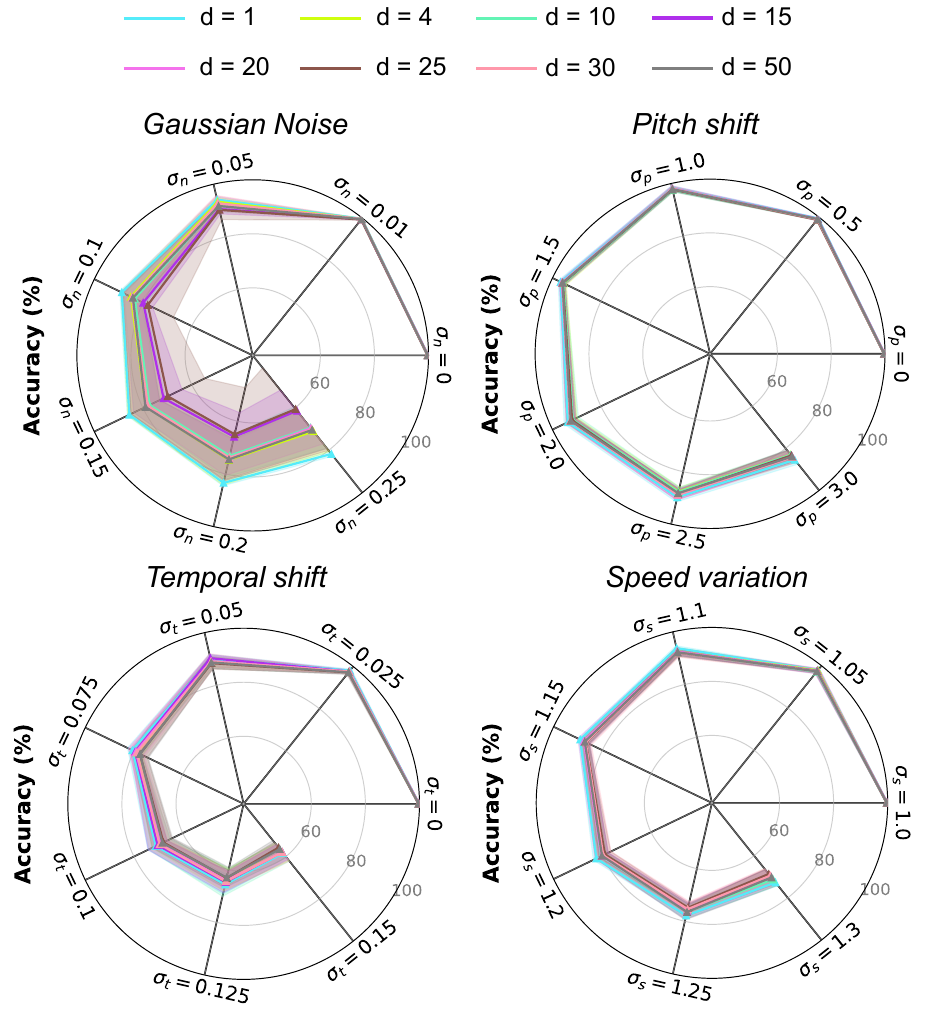}
    \caption{Accuracy of the QNN-Random model across varying circuit depths (\(d \in [1, 4, 10, 15, 20, 25, 30, 50]\)) under different corruption types on TESS dataset. The model achieves its highest accuracy and robustness at a shallow depth of \(d = 1\), with performance degrading non-linearly as the depth increases.}
    \vspace{-20pt}
    \label{fig:Random all layer}
\end{figure}
Fig.~\ref{fig:Random all layer} further illustrates the accuracy of the QNN-Random model under four types of input corruptions on TESS dataset, while the corresponding results for QNN-Basic and QNN-Strongly on TESS, AVFAD datasets are provided in the Supplementary Material. The results indicate that circuit depth has an impact on model robustness, particularly under Gaussian noise. For all four corruption types, the QNN-Random model achieves its highest accuracy and robustness at a shallow depth of \(d = 1\). As the depth increases, performance degrades in a non-linear manner, suggesting that deeper circuits do not necessarily enhance generalization. Similar non-linear degradation trends are observed for QNN-Basic and QNN-Strongly; however, their optimal depths vary across different corruption types, as detailed in the Supplementary Material. 

Overall, varying the circuit depth in QNN-Random and QNN-Strongly has only a minor influence on robustness to pitch shift, temporal shift, and speed variation, whereas QNN-Basic exhibits greater sensitivity to depth changes. In contrast, all three QNN variants show a pronounced dependence on depth when subjected to Gaussian noise, underscoring the critical role of circuit complexity in maintaining noise robustness.
\vspace{-10pt}
\subsection{Convergence Analysis}
\label{subsec: convergence}
\begin{figure}[ht!]
    \centering
    \begin{subfigure}[b]{0.24\textwidth}
        \centering
        \includegraphics[width=\textwidth]{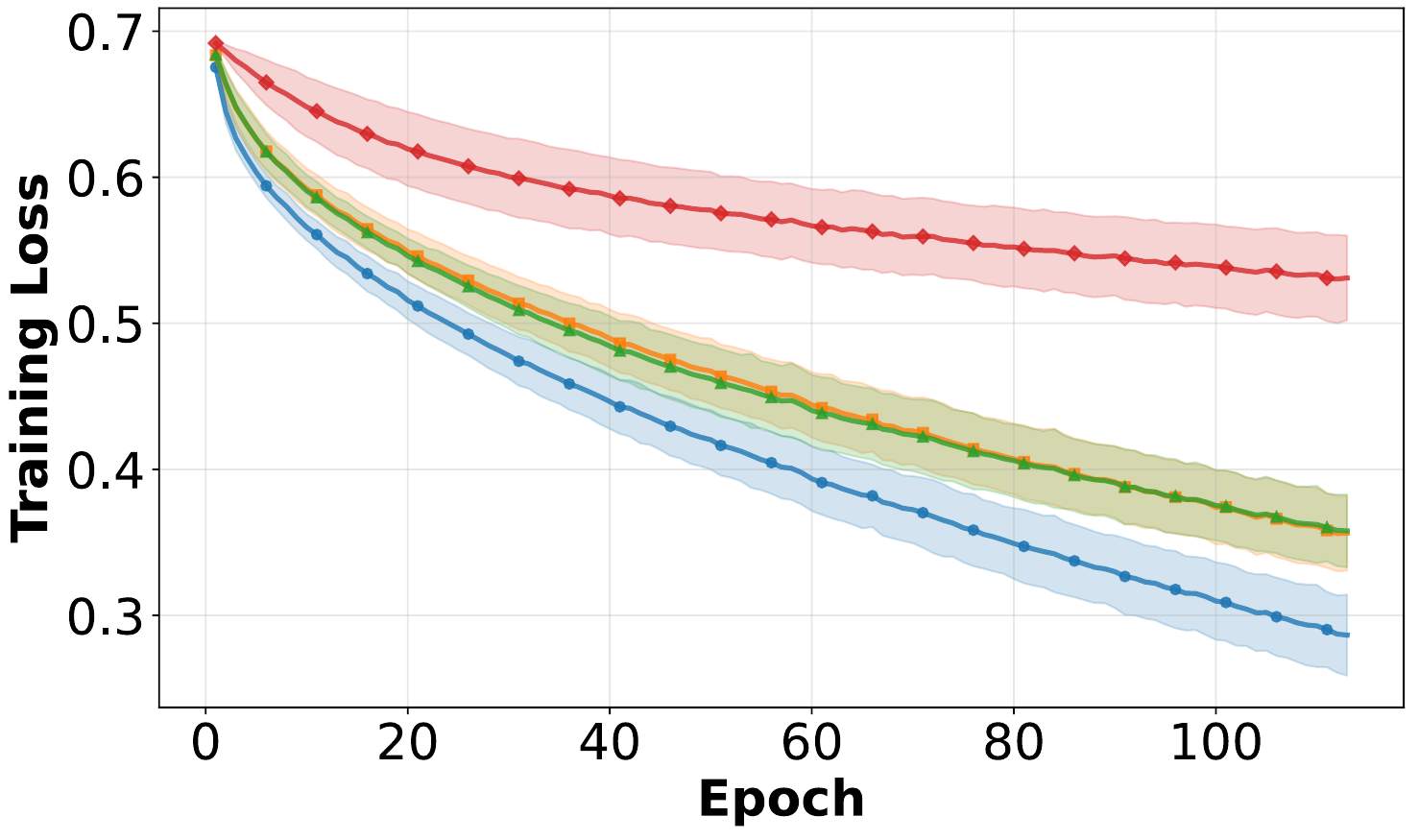}
        \caption{Training loss.}
        \label{fig: avfad_training_loss}
    \end{subfigure}
    \begin{subfigure}[b]{0.24\textwidth}
        \centering
        \includegraphics[width=\textwidth]{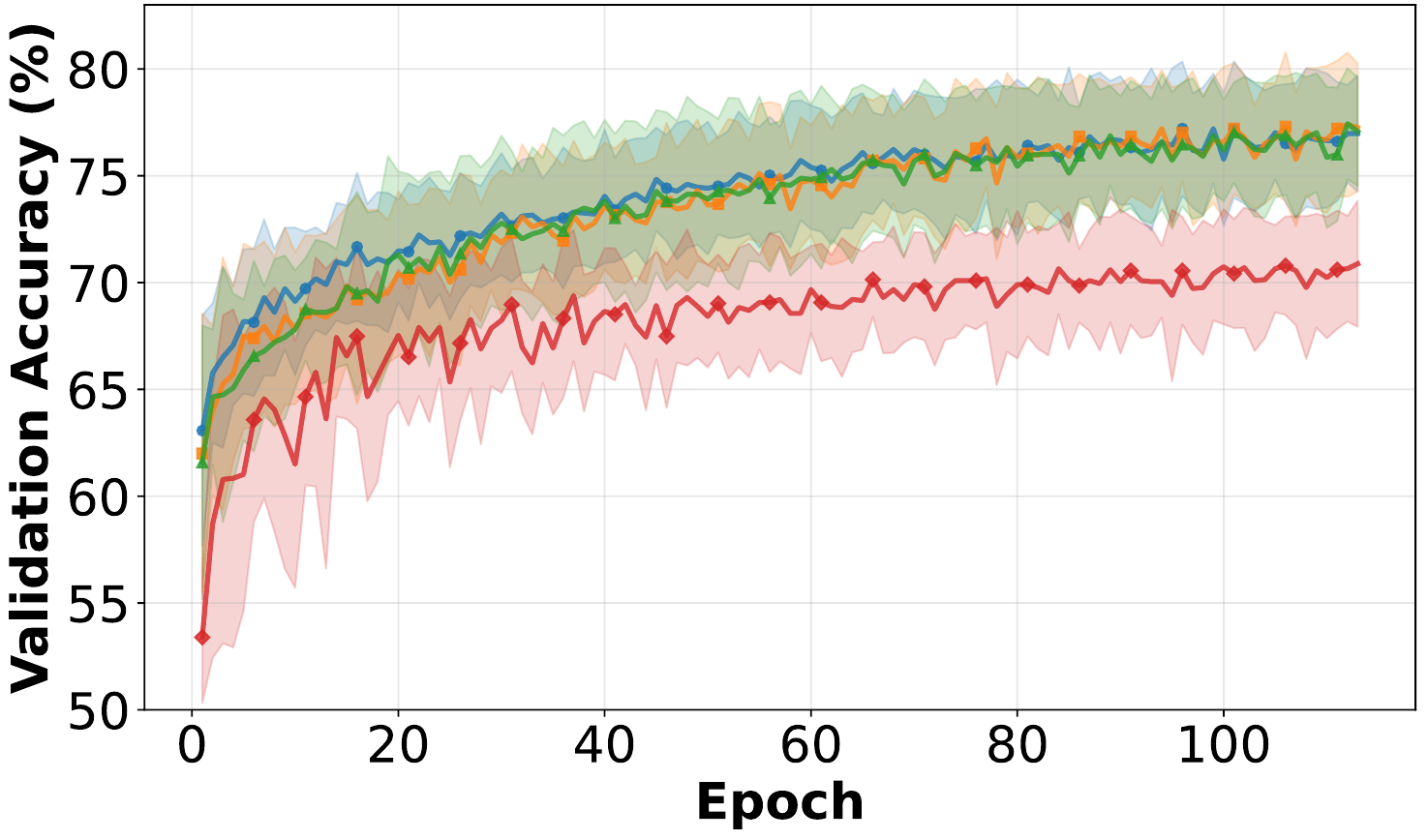}
        \caption{Validation accuracy.}
        \label{fig: avfad_validation_acc}
    \end{subfigure}
    \caption{Training loss and validation accuracy on the AVFAD dataset. The corresponding circuit depths ($d$) for QNN-Basic, QNN-Strongly, and QNN-Random are summarized in Table~\ref{tab:depth_circuit}.}
    \label{fig: Avfad_training_process}
    \vspace{-10pt}
\end{figure}

\begin{figure}[ht!]
    \centering
    \begin{subfigure}[b]{0.24\textwidth}
        \centering
        \includegraphics[width=\textwidth]{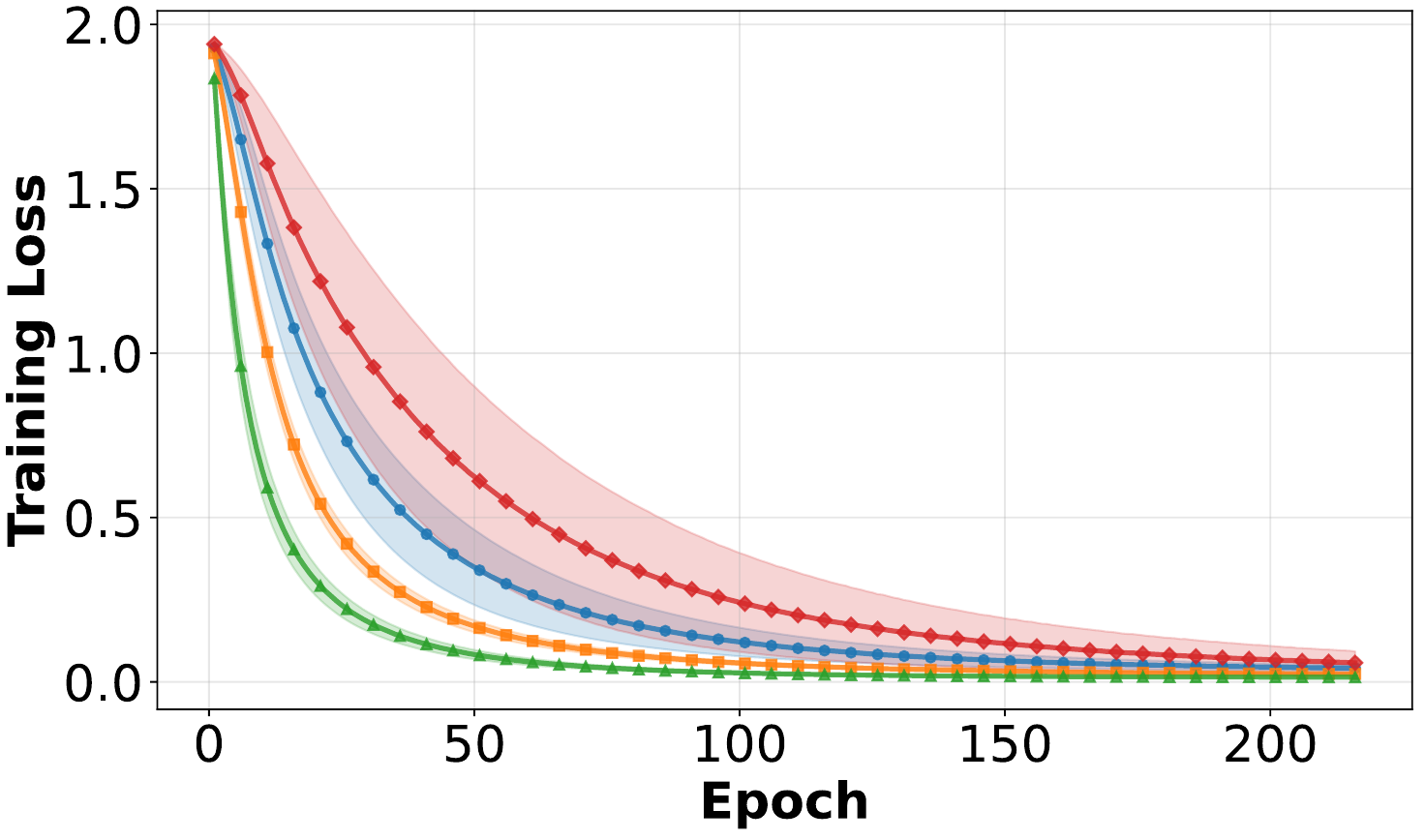}
        \caption{Training loss.}
        \label{fig: tess_training_loss}
    \end{subfigure}
    \begin{subfigure}[b]{0.24\textwidth}
        \centering
        \includegraphics[width=\textwidth]{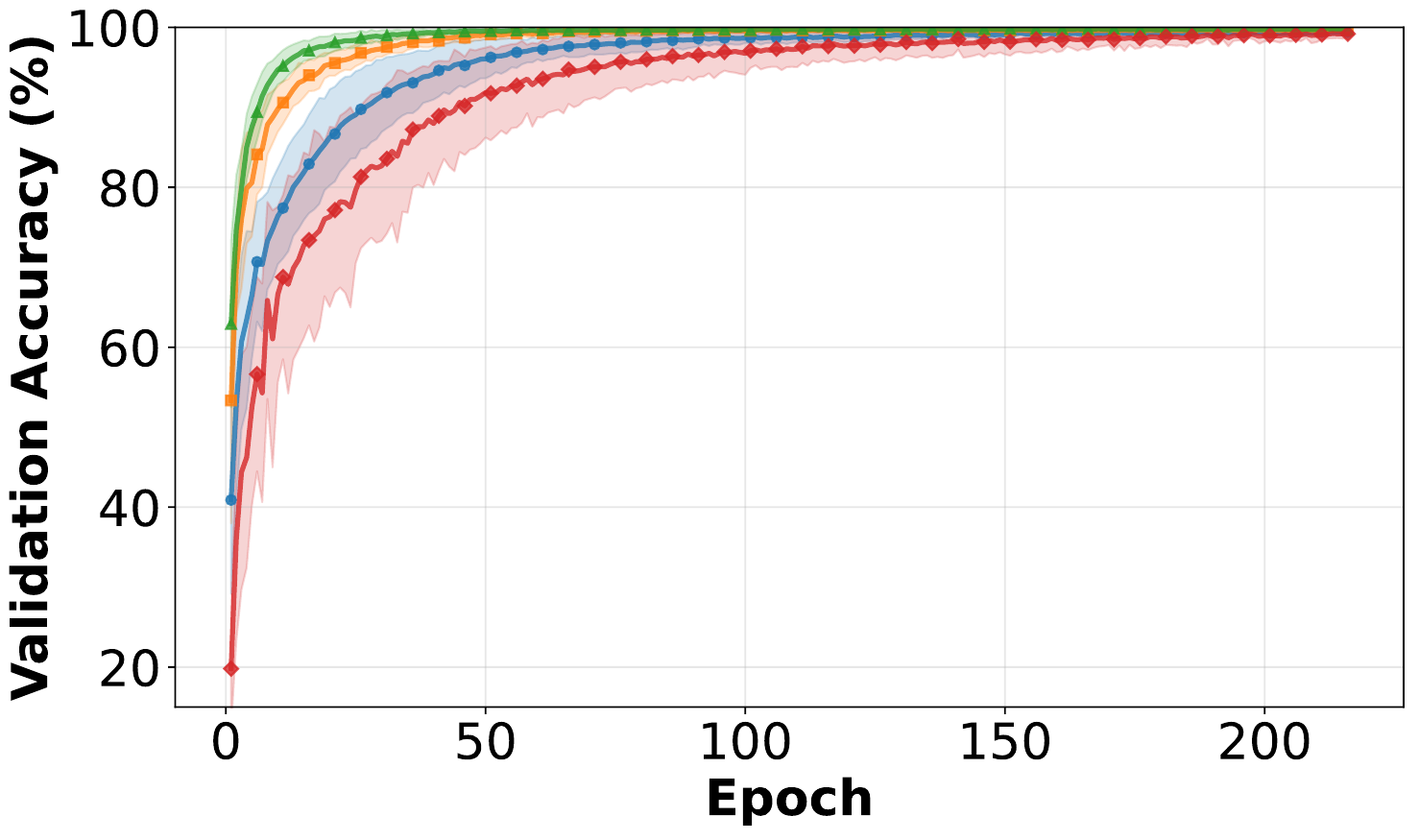}
        \caption{Validation accuracy.}
        \label{fig: tess_validation_acc}
    \end{subfigure}
    \caption{Training loss and validation accuracy on the TESS dataset. The corresponding circuit depths ($d$) for QNN-Basic, QNN-Strongly, and QNN-Random are summarized in Table~\ref{tab:depth_circuit}.}
    \label{fig: tess_training_process}
    \vspace{-10pt}
\end{figure}

Figures \ref{fig: Avfad_training_process} and \ref{fig: tess_training_process} present the training loss and validation accuracy per epoch on the AVFAD and TESS datasets, with values reported as the mean and standard deviation across 10 random seeds. Early stopping was applied during training, and the minimal number of epochs is shown to highlight convergence behaviour. As expected, the training loss decreases and validation accuracy increases as the number of epochs grows.
On the AVFAD dataset, CNN-Base consistently exhibits lower validation accuracy and higher loss compared to QNNs. Across both datasets, QNNs converge more rapidly than CNN-Base, with the effect being particularly pronounced on TESS. For example, QNN-Random converges within approximately 30 epochs, whereas CNN-Base requires nearly 200 epochs. On AVFAD, QNN-Basic achieves the fastest convergence, while on TESS, QNN-Random converges most quickly, which is consistent with the robustness findings reported in Section~\ref{subsec:Robustness Analysis}. Furthermore, CNN-Base shows a larger standard deviation on TESS, indicating greater variability across seeds compared to QNNs.

In summary, QNNs converge substantially faster than CNN-Base on both datasets, a property that is advantageous in real-world scenarios where faster training can reduce computational cost and improve deployment efficiency.

\section{Conclusion}
\label{sec: conclusion}
This study systematically evaluated the robustness of multiple QNN models against classical CNNs across diverse acoustic corruptions using voice-pathology (AVFAD) and speech-emotion (TESS) datasets. Across corruption types, QNNs consistently surpassed the CNN-Base under pitch shift, temporal shift, and speed variation (up to 22\% advantage in CE/RCE under severe temporal shift), while the CNN-Base maintained an advantage under Gaussian noise, highlighting complementary noise sensitivities between quantum and classical models. Dataset characteristics also influenced robustness: QNNs sustained competitive performance on AVFAD’s irregular and noisy speech, whereas on the cleaner TESS dataset Gaussian noise disproportionately degraded QNNs, plausibly amplifying random fluctuations in the quanvolutional layer. Moreover, QNNs also reach competitive and exceed the robustness of two advanced classical architectures (ResNet-18 and VGG-16) while remaining smaller number of parameter. Emotion-wise analysis revealed that \textit{fear} remained the most robust class (approximately 80–90\% accuracy under severe temporal corruptions), whereas \textit{neutral} and \textit{happy} were most vulnerable to Gaussian noise and temporal shift, respectively. Among the three quantum circuits, QNN-Basic demonstrated the highest overall robustness, particularly on AVFAD, achieving the lowest mCE and RmCE values. QNN-Random exhibited superior generalization on TESS, outperforming the CNN-Base under most corruption types except Gaussian noise, whereas QNN-Strongly showed reduced resilience, suggesting that excessive entanglement depth can degrade robustness. Moreover, the depth of the quantum circuit impacts on the performance of QNNs, especially under Gaussian noise. Finally, training dynamics demonstrated a marked convergence advantage, with QNNs stabilizing within roughly 30 epochs compared to the CNN-Base requiring nearly 200 epochs. Collectively, these findings indicate that QNNs not only converge faster but also exhibit intrinsic robustness to structural and temporal corruptions, reinforcing their potential as resilient architectures for speech-based QML. Our analysis is conducted with software-simulated quantum layers, two speech datasets, and non-reverberant corruption conditions; we do not model room acoustics, channel effects, overlapping speakers, or quantum hardware noise.

In future work, we will evaluate QNN robustness under more realistic acoustic conditions such as reverberation, channel distortions, and overlapping speech, and we will explicitly integrate quantum hardware noise into training and inference. Additionally, we plan to investigate the performance benefits of trainable quanvolutional layers and conduct cross-corpus evaluations by incorporating additional datasets to assess model generalization across diverse speech corpora.
\vspace{-10pt}
\section*{Acknowledgments}
This research is supported by Network Sensing \& Biomedical Engineering (NSBE) Research Lab, School of Engineering, Deakin University. 

\bibliographystyle{IEEEtran}
\bibliography{References}

\vfill

\end{document}